\documentclass[12pt]{iopart}
\begin{document}
\title[Matter with dilaton charge and evolution of the Universe]
{Matter with dilaton charge in Weyl--Cartan spacetime and evolution of the
Universe }

\author{Olga V Babourova\dag\footnote[3]{babourova@mail.ru} and
Boris N Frolov\ddag\footnote[4]{frolovbn@mail.ru}}

\address{\dag\ Department of Theoretical Physics, Faculty of Physics,
Moscow State University, Vorobjovy Gory, Moscow 119992, Russian Federation}

\address{\ddag\ Department of Physics, Faculty of Mathematics, Moscow State
Pedagogical University, Krasnoprudnaya 14, Moscow 107140, Russian Federation}

\begin{abstract}
The perfect dilaton-spin fluid (as a model of the dilaton matter, the
particles of which are endowed with intrinsic spin and dilaton charge) is
considered as the source of the gravitational field in a Weyl--Cartan
spacetime.  The variational theory of such fluid is constructed  and the
dilaton-spin fluid energy-momentum tensor is obtained. The variational
formalism of the gravitational field in a Weyl--Cartan spacetime is
developed in the exterior form language. The homogeneous and isotropic
Universe filled with the dilaton matter as the dark matter is considered
and one of the field equations is represented as the Einstein-like equation
which leads to the modified Friedmann--Lema\^{\i}tre equation. From this
equation the absence of the initial singularity in the cosmological
solution follows. Also the existence of two points of inflection of the
scale factor function is established, the first of which corresponds to the
early stage of the Universe and the second one corresponds to the modern
era when the expansion with deceleration is replaced by the expansion with
acceleration.  The possible equations of state for the self-interacting
cold dark matter are found on the basis of the modern observational data.
The inflation-like solution is obtained.
\end{abstract}


\newpage

\section{Introduction}
\setcounter{equation}{0}

The basis concept of the modern fundamental physics consists in the
preposition that the spacetime geometrical structure is compatible with the
properties of matter filling the spacetime. It means that the matter dynamics
determines the metric and the connection of the spacetime manifold and
in turn is determined by the spacetime geometric properties. Therefore the
possible deviation from the geometrical structure of the General Relativity
spacetime should be stipulated by the existence of matter with unusual
properties, which fills spacetime, generates its structure and interacts
with it. As examples of such matter there were considered the perfect media
with intrinsic degrees of freedom, such as the perfect fluid with spin and
non-Abelian colour charge \cite{bfdan}, the perfect hypermomentum fluid (see
\cite{bfhyp} and the references therein), the perfect dilaton-spin fluid
\cite{bfmod1,bfmod2,bfGrC}. All these fluids are the generalization of the
perfect Weyssenhoff--Raabe spin fluid \cite{WR}.

The modern observations \cite{nabl} lead to the conclusion
about the existence of dark (non\-lu\-mi\-nous) matter with the density
exceeding by one order of magnitude the density of baryonic matter, from
which stars and luminous components of galaxies are formed. It is the dark
matter interacting with the equal by order of magnitude positive vacuum
energy or with quintessence \cite{Os-Perl,St} that realizes the dynamics
of the Universe in modern era. Another important consequence of the modern
observations is the understanding of the fact that the end of the Friedmann
era occurs when the expansion with deceleration is succeeded by the
expansion with acceleration, the transition to the unrestrained exponential
expansion being possible.

As dark matter we propose to consider the dilaton matter, the model of which
is realized as the perfect dilaton-spin fluid. Every particle of such fluid
is endowed with spin and dilaton charge. This type of matter generates in
spacetime the Weyl--Cartan geometrical structure with curvature, torsion and
nonmetricity of the Weyl type and interacts with it. Such hypothetical type
of matter was proposed in \cite{bfmod1,bfmod2}, where the variational
theory of perfect fluid of this type has been constructed, and the
peculiarities of motion of the particles with spin and dilaton charge in
a Weyl--Cartan spacetime have been considered.

The importance of considering matter endowed with the dilaton charge is
based on the fact that some variants of low-energy effective string theory
are reduced to the theory of interacting metric and scalar dilaton field
\cite{Gr-SW} represented by the gradient part of the Weyl connection. It
should be pointed out that this gradient part was successfully used in
\cite{Perv1} for explanation some effects of the modern cosmology.
Therefore the dilaton gravity is one of the attractive approaches to the
modern gravitational theory.

The Weyl--Cartan cosmology was considered in \cite{Min1,Ob-Hel,Tres}
as a particular case of the metric--affine theory of gravitation. In
our approach we consider the Weyl--Cartan spacetime geometry from the very
beginning, the Weyl's restriction on a nonmetricity 2-form being imposed
with a Lagrange multiplier. This approach in the exterior form language
was developed in \cite{Pont,bfGrC}. These two approaches are
not identical and yield the different field equations in general, as it will
be pointed out in section 4.

Our paper is organized as follows.

In section 2, the variational theory of the perfect dilaton-spin fluid
is constructed in a Weyl--Cartan spacetime with torsion and
nonmetricity with the help of the exterior form formalism. In section 3,
the dilaton-spin fluid energy-momentum tensor and dilaton-spin current
are obtained, and the energy conservation law along a streamline of the fluid
is derived as a consequence of the generalized hydrodynamic Euler-type
equation of the perfect fluid motion. In section 4, the variational formalism
of the gravitational field in a Weyl--Cartan spacetime is developed and
two gravitational field equations, $\Gamma$- and $\theta$-equations, are
derived. In section 5, the general solution of the field $\Gamma$-equation is
obtained. In section 6, the homogeneous and isotropic Universe is considered
and the field $\theta$-equation is represented as the Einstein-like equation
which leads to the modified Friedmann--Lema\^{\i}tre (FL) equation. From this
equation it follows (section 7) that, firstly, under the certain conditions
on the parameters of the gravitational Lagrangian this equation has a
nonsingular solution and, secondly, two points of inflection of the scale
factor function exist.  The first one corresponds to the early stage of the
Universe and the second one corresponds to the modern era when the
expansion with deceleration is replaced by the expansion with acceleration.
Here the possible equations of state for the self-interacting cold dark
matter are found on the basis of the modern observational data. The
inflation-like solution is obtained for the superrigid equation of state of
the dilaton matter at the very early stage of the evolution of the Universe.
Section 8 is devoted to the discussion.

    Some facts of a Weyl--Cartan space and the notations used are stated in
Appendix A, in particular, the signature of the metric is assumed to be
$(+,+,+,-)$. In Appendices B and C the results of the intermediate
calculations are stated.  Throughout the paper the conventions $c = 1$,
$\hbar = 1$ are used.

\section{The variational theory of the dilaton-spin fluid}
\setcounter{equation}{0}

Let us develop the variational theory of the perfect dilaton-spin fluid in a
Weyl--Cartan space. We shall describe the additional degrees of
freedom, which a fluid element possesses, as it is accepted in the theory of
the Weyssenhoff--Raabe fluid, with the help of the material frame attached
to every fluid element and consisting of four vectors $\vec{l}_{p}$
($p=1,2,3,4$) and inverse to them vectors $\vec{l}^{q}$. The vectors $\vec{l}
_{p}$ are called directors. Three of the directors ($p=1,2,3$) are
space-like and the fourth ($p=4$) is time-like. In the exterior form
language the material frame of the directors turns into the coframe of
1-forms $l^{p}$, which have dual 3-forms $l_{q}$, the constraint
$l^{p}\wedge l_{q}=\delta _{q}^{p}\eta $ being fulfilled. This constraint
means that $l_{\alpha }^{p}l_{p}^{\beta }=\delta _{\alpha }^{\beta }$. Each
fluid element possesses a 4-velocity vector $\vec{u}=u^{\alpha }\vec{e}
_{\alpha }$ which corresponds to a flow 3-form $u:=\vec{u}\rfloor \eta
=u^{\alpha }\eta _{\alpha }$ and a velocity 1-form $\ast u=u_{\alpha }\theta
^{\alpha }=\breve{g}(\vec{u},\cdots )$ with $\ast u\wedge u=-\eta $ that
means the usual condition $\breve{g}(\vec{u},\vec{u})=-1$.

In case of the dilaton-spin fluid the spin dynamical variable of the
Weyssenhoff--Raabe fluid is generalized and becomes the new dynamical
variable $J^{p}\!_{q}$ named the dilaton-spin tensor \cite{bfmod1}:
\[
J^{p}\!_{q} = S^{p}\!_{q} + \frac{1}{4} \delta^{p}_{q}J\;, \qquad
S_{pq} = J_{[pq]}\;, \qquad J = J^{p}\!_{p}\; .
\]
It is well-known due to Frenkel theory of the rotating electron \cite{Fren}
that the spin tensor $S^{p}\!_{q}$ of a particle is spacelike in its nature
that is the fact of fundamental physical meaning. This leads to the
classical Frenkel condition, which can be expressed in
two equivalent forms: $S^{p}\!_{q} u^{q} =~0$, $u_{p}S^{p}\!_{q} = 0$ or in
the exterior form language, $S^{p}\!_{q} l_{p}\wedge*\!u= 0$, $S^{p}\!_{q}
l^{q} \wedge u = 0$. It should be mentioned that the Frenkel condition
appears to be a consequence of the generalized conformal invariance of the
Weyssenhoff perfect spin fluid Lagrangian \cite{Frol}.

It is important that only the first term of $J^{p}\!_{q}$ (the spin tensor)
obeys to the Frenkel condition \cite{bfmod1}. The second term is
proportional to the specific (per particle) dilaton charge $J$ of the
fluid element. The existence of the dilaton charge is the consequence of the
extension of the Poincar\'{e} symmetry (with the spin tensor as the
dynamical invariant) to the Poincar\'{e}--Weyl symmetry with the
dilaton-spin tensor as the dynamical invariant.

The measure of intrinsic motion contained in a fluid element is the quantity
$\Omega^{q}\!_{p}$ which generalizes the intrinsic `angular velocity' of the
Weyssenhoff spin fluid theory,
\[
\Omega^{q}\!_{p} \eta = u \wedge l^{q}_{\alpha} {\rm D} l^{\alpha}_{p} \; ,
\qquad {\rm D} l^{\alpha}_{p} = \rmd l^{\alpha}_{p} +
\Gamma^{\alpha}\!_{\beta} l^{\beta}_{p} \; .
\]
An element of the perfect dilaton-spin fluid possesses the additional
intrinsic `kinetic' energy density 4-form,
\[
E = \frac{1}{2} n J^{p}\!_{q}\Omega^{q}\!_{p}\eta = \frac{1}{2} n
S^{p}\!_{q} u\wedge l^{q}_{\alpha} {\rm D} l^{\alpha}_{p} + \frac{1}{8} n J
u\wedge l^{p}_{\alpha} {\rm D} l^{\alpha}_{p}\; ,
\]
where $n$ is the fluid particles concentration.

The fluid element represents the quasi-closed statistical subsystem (with
sufficiently large number of particles), the properties of which coincide
with the macroscopic pro\-per\-ti\-es of the fluid. The internal energy
density of the fluid $\varepsilon$ depends on the extensive (additive)
thermodynamic parameters: the fluid particles concentration $n$ and the
specific (per particle) entropy $s$ of the fluid in the rest frame of
reference, the specific spin tensor $S^{p}\!_{q}$ and the specific dilaton
charge $J$. The energy density $\varepsilon$ obeys to the first
thermodynamic principle,
\begin{equation}
\rmd\varepsilon(n, s, S^{p}\!_{q}, J) = \frac{\varepsilon + p}{n} \rmd n +
n T \rmd s + \frac{\partial\varepsilon}{\partial S^{p}\!_{q}} \rmd S^{p}\!_{q}
+ \frac {\partial \varepsilon}{\partial J} \rmd J \;,  \label{eq:d13}
\end{equation}
where $p$ is the hydrodynamic fluid pressure, the fluid particles number
conservation law $\rmd (nu) = 0$ being taken into account. The entropy
conservation law $\rmd (nsu) = 0$ is also fulfilled along the streamline
of the fluid.

The Lagrangian density 4-form of the perfect dilaton-spin fluid can be
constructed from the quantities $\varepsilon$ and $E$ with regard to the
constraints which should be introduced into the Lagrangian density by means
of the Lagrange multipliers $\varphi$, $\tau$, $\chi$, $\chi^{q}$ and
$\zeta_{p}$,
\begin{eqnarray}
\fl {\cal L}_{{\rm{fluid}}} = - \varepsilon (n, s, S^{p}\!_{q}, J) \eta +
\frac{1}{2}n S^{p}\!_{q} u\wedge l^{q}_{\alpha} {\rm D} l^{\alpha}_{p} +
\frac{1}{8} n J u\wedge l^{p}_{\alpha} {\rm D} l^{\alpha}_{p} + n u \wedge
\rmd \varphi \nonumber \\
+ n \tau u\wedge \rmd s + n \chi (*\!u \wedge u
+ \eta) + n \chi^{q} S^{p}\!_{q} l_{p}\wedge *\! u +
n\zeta_{p}S^{p}\!_{q}l^{q}\wedge u\; . \label{eq:20}
\end{eqnarray}

The fluid motion equations and the evolution equation of the dilaton-spin
tensor are derived by the variation of (\ref{eq:20}) with respect to the
independent variables $n$, $s$, $S^{p}\!_{q}$, $J$, $u$, $l^{q}$ and the
Lagrange multipliers, the thermodynamic principle (\ref{eq:d13}) being taken
into account and master-formula (\ref{eq:master}) (see Appendix B) being
used. We shall consider the 1-form $l^{q}$ as an independent variable and
the 3-form $l_{p}$ as a function of $l^{q}$. One can verify that the
Lagrangian density 4-form (\ref{eq:20}) is proportional to the hydrodynamic
fluid pressure, ${\cal L}_{{\rm fluid}} = p \eta$.

The variation with respect to the material coframe $l^q$ yields the motion
equations of the directors, which lead to the dilaton charge conservation
law $\dot{J} = 0$ and to the evolution equation of the spin tensor,
\begin{equation}
\dot{S}^{\alpha}\!_{\beta} + \dot {S}^{\alpha}\!_{\gamma}u^{\gamma}
u_{\beta} + \dot {S}^{\gamma}\!_{\beta}u_{\gamma}u^{\alpha} = 0\; ,
\label{eq:d33}
\end{equation}
where the `dot' notation for the tensor object $\Phi$ is introduced, $\dot
{\Phi}^{\alpha}\!_{\beta} = *\!(u\wedge {\rm D}\Phi^{\alpha}\!_ {\beta})$.
The equation (\ref{eq:d33}) generalizes the evolution equation of the spin
tensor of the Weyssenhoff fluid theory to a Weyl--Cartan space.
With the help of the projection tensor $\Pi^{\alpha}_{\sigma}$ these two
equations can be represented in the equivalent form \cite{bfmod1},
$\Pi^{\alpha}_{\sigma} \Pi^{\rho}_{\beta}\dot{J}^{\sigma}\!_{\rho} = 0$.

\section{The energy-momentum 3-form and the hydrodynamic Euler equation}
\setcounter{equation}{0}

By means of the variational derivatives of the matter Lagrangian density
(\ref{eq:20}) one can derive the external matter currents which are the
sources of the gravitational field. In case of the perfect dilaton-spin
fluid the matter currents are the canonical energy-momentum 3-form $%
\Sigma_{\sigma}$, the metric stress-energy 4-form $\sigma^{\alpha\beta}$,
the dilaton-spin momentum 3-form ${\cal J}^{\alpha}\!_{\beta}$, derived in
\cite{bfmod1}.

The variational derivative of the Lagrangian density (\ref{eq:20}) with
respect to $\theta^{\sigma}$ yields the canonical energy-momentum 3-form,
\begin{equation}
\Sigma_{\sigma} = \frac{\delta {\cal L}_{{\rm{fluid}}}}{\delta \theta^{\sigma}}
= p\eta_{\sigma} + (\varepsilon + p)u_{\sigma} u + n \dot{S}_{\sigma}\!_{\rho}
u^{\rho} u\; .  \label{eq:d37}
\end{equation}
Here the Frenkel condition, the dilaton charge conservation law
$\dot{J} = 0$ and the evolution equation of the spin tensor (\ref{eq:d33})
have been used. In case of the dilaton-spin fluid the energy density
$\varepsilon$ in (\ref{eq:d37}) contains the energy density of the
dilaton interaction of the fluid.

The metric stress-energy 4-form can be derived in the same way,
\begin{eqnarray}
&& \sigma^{\alpha\beta} = 2\frac{\delta {\cal L}_{{\rm{fluid}}}}{\delta
g_{\alpha\beta}} = T^{\alpha\beta} \eta \; ,  \nonumber \\
&& T^{\alpha\beta} = p g^{\alpha\beta} + (\varepsilon +
p)u^{\alpha}u^{\beta} + n \dot{S}^{(\alpha}\!_{\gamma}u^{\beta )} u^{\gamma}
\;.  \label{eq:d353}
\end{eqnarray}

The dilaton-spin momentum 3-form can be obtained in the following way,
\begin{equation}
{\cal J}^{\alpha }\!_{\beta }=-\frac{\delta {\cal L}_{\rm{fluid}}}{\delta
\Gamma^{\beta }\!_{\alpha }}=\frac{1}{2}n\left( S^{\alpha }\!_{\beta }
+ \frac{1}{4}J\delta _{\beta }^{\alpha }\right) u =  {\cal S}_{\alpha\beta}
+ \frac{1}{4}{\cal J} \delta^\alpha_\beta \;.  \label{eq:d38}
\end{equation}

In a Weyl--Cartan space the matter Lagrangian obeys the
diffeomorphism inva\-ri\-ance, the local Lorentz invariance and the local
scale invariance that lead to the corres\-pon\-d\-ing Noether identities
\cite{bfmod2}:
\begin{eqnarray}
&& {\rm D}\Sigma_{\sigma} = (\vec{e}_{\sigma}\rfloor {\cal T}
^{\alpha})\wedge \Sigma_{\alpha} - (\vec{e}_{\sigma}\rfloor {\cal R}
^{\alpha}\!_{\beta})\wedge {\cal J}^{\beta}\!_{\alpha} - \frac{1}{8}(\vec{e}
_{\sigma}\rfloor {\cal Q}) \sigma^{\alpha}\!_{\alpha}\; , \label{eq:dzak1}\\
&&\left ({\rm D} +\frac{1}{4}{\cal Q}\right )\wedge {\cal S}_{\alpha\beta}
= \theta_{ [\alpha} \wedge \Sigma_{\beta ]} \; ,  \label{eq:d39} \\
&& {\rm D}{\cal J} = \theta^{\alpha} \wedge \Sigma_{\alpha} -
\sigma^{\alpha}\!_{\alpha}\; .  \label{eq:d391}
\end{eqnarray}
The Noether identity (\ref{eq:dzak1}) represents the quasiconservation law
for the canonical matter energy-momentum 3-form. This identity and also the
identities (\ref{eq:d39}) and (\ref{eq:d391}) are fulfilled, if the
equations of matter motion are valid, and therefore they represent in
essence another form of the matter motion equations.

If one introduces a specific (per particle) dynamical momentum of a fluid
element,
\[
\pi_{\sigma}\eta = -\frac{1}{n}*\!u\wedge \Sigma_{\sigma}\; , \qquad
\pi_{\sigma} = \frac{\varepsilon}{n}u_{\sigma} - S_{\sigma\rho}\vec{u}
\rfloor {\rm D} u^{\rho}\;,
\]
then the canonical energy-momentum 3-form reads,
\begin{equation}
\Sigma_{\sigma} = p\eta_{\sigma} + n \left (\pi_{\sigma} + \frac{p}{n}
u_{\sigma}\right ) u\; .  \label{eq:dsig}
\end{equation}

Substituting (\ref{eq:dsig}), (\ref{eq:d353}) and (\ref{eq:d38}) into (\ref
{eq:dzak1}), one obtains after some algebra the equation of motion of the
perfect dilaton-spin fluid in the form of the generalized hydro\-dy\-na\-mic
Euler-type equation of the perfect fluid \cite{bfmod2},
\begin{eqnarray}
\fl u\wedge {\rm D} \left (\pi_{\sigma} + \frac{p}{n}u_{\sigma}\right ) =
\frac{1}{n}\eta\vec{e}_{\sigma}\rfloor {\rm D}p - \frac{1}{8n}\eta
(\varepsilon + p) Q_{\sigma} - (\vec{e}_{\sigma}\rfloor{\cal T}^{\alpha})
\wedge \left (\pi_{\alpha} + \frac{p}{n}u_{\alpha}\right) u \nonumber \\
- \frac{1}{2} (\vec{e}_{\sigma}\rfloor {\cal R}^{\alpha\beta})\wedge
S_{\alpha\beta}u + \frac{1}{8} (\vec{e}_{\sigma}\rfloor
{\cal R}^{\alpha}\!_{\alpha})\wedge Ju \; .  \label{eq:deuler}
\end{eqnarray}

Let us evaluate the component of the equation (\ref{eq:deuler}) along the
4-velocity by contracting one with $u^{\sigma}$. After some algebra we get
the energy conservation law along a streamline of the fluid \cite{bfmod2},
\begin{equation}
\rmd\varepsilon = \frac{\varepsilon + p}{n} \rmd n \; .  \label{eq:cons}
\end{equation}

\section{Variational formalism in a Weyl--Cartan space}
\setcounter{equation}{0}

We represent the total Lagrangian density 4-form of the theory as follows
\begin{equation}
{\cal L} = {\cal L}_{{\rm{grav}}} + {\cal L}_{\rm{fluid}}\; ,  \label{eq:tot}
\end{equation}
where the gravitational field Lagrangian density 4-form reads,
\begin{eqnarray}
\fl {\cal L}_{\rm{grav}} =2f_{0}\biggl (\frac{1}{2}{\cal R}^{\alpha}\!_{\beta}
\wedge \eta_\alpha\!^\beta  - \Lambda \eta + \frac{1}{4}\lambda \,
{\cal R}^\alpha\!_\alpha \wedge *{\cal R}^\beta\!_\beta + \varrho_1 \,
{\cal T}^{\alpha} \wedge *{\cal T}_{\alpha}  \nonumber \\
+ \varrho_2 \,({\cal T}^{\alpha} \wedge
{\theta}_{\beta}) \wedge *({\cal T}^{\beta}\wedge{\theta}_{\alpha}) +
\varrho_3 \,({\cal T}^{\alpha}\wedge {\theta}_{\alpha})\wedge * ({\cal T}
^{\beta}\wedge {\theta}_{\beta}) \nonumber \\
+ \xi \,{\cal Q}\wedge * {\cal Q} + \zeta \,{\cal Q}\wedge \theta^\alpha
\wedge *{\cal T}_\alpha\biggr ) + \Lambda^{\alpha\beta}\wedge \left ({\cal Q}
_{\alpha\beta} - \frac{1}{4}g_{\alpha\beta}{\cal Q}\right )\;. \label{eq:tlag}
\end{eqnarray}
Here $f_0 = 1/(2\mbox{\ae})$ ($\mbox{\ae} = 8\pi G $), $\Lambda$ is the
cosmological constant, $\lambda$, $\varrho_1$, $\varrho_2$, $\varrho_3$,
$\xi$, $\zeta$ are the coupling constants, and $\Lambda^{\alpha\beta}$ is
the Lagrange multiplier 3-form with the evident properties,
\begin{equation}
\Lambda^{\alpha\beta} = \Lambda^{\beta\alpha} \;, \qquad
\Lambda^\gamma\!_\gamma = 0 \;,  \label{eq:svlam}
\end{equation}
which are the consequences of the Weyl's condition (\ref{eq:Q}).

In (\ref{eq:tlag}) the first term is the linear Gilbert--Einstein Lagrangian
generalized to a Weyl--Cartan space. The second term is the Weyl quadratic
Lagrangian, which is the square of the Weyl segmental curvature 2-form
(see (\ref{eq:Grusl}) and (\ref{eq:Bian})),
\begin{equation}
{\cal R}^{\alpha}\!_{\alpha} = \rmd \Gamma^{\alpha}\!_{\alpha} =
\frac{1}{2}\rmd{\cal Q} \;.  \label{eq:hom}
\end{equation}
Here the Weyl 1-form ${\cal Q}$, in contrast to the classical Weyl theory,
represents the gauge field, which does not relate to an electromagnetic field,
that has been firstly pointed out by Utiyama \cite{Ut}. The field of the
Weyl 1-form ${\cal Q}$ we shall call a dilaton field. The term with the
coupling constant $\zeta$ represents the contact interaction of the dilaton
field with the torsion that can occur in a Weyl--Cartan space.

The gravitational field equations in a Weyl--Cartan spacetime can be obtained
by the variational procedure of the first order. Let us vary the Lagrangian
(\ref{eq:tot}) with respect to the connection 1-form $\Gamma^{\alpha}\!_{\beta}$
($\Gamma$-equation) and to the basis 1-form $\theta^\alpha$ ($\theta$-equation)
indepen\-den\-t\-ly, the constraints on the connection 1-form in
a Weyl--Cartan space being satis\-fi\-ed by means of the Lagrange
multiplier 3-form $\Lambda^{\alpha \beta}$.

The including into the Lagrangian density 4-form the term with the Lagrange
mul\-ti\-pli\-er $\Lambda ^{\alpha \beta }$ means that the theory is
considered in a Weyl--Cartan spacetime from the very beginning
\cite{Pont,bfGrC}. Another variational approach has been developed in
\cite{Ob-Hel,Tres} where the field equations in a Weyl--Cartan spacetime
have been obtained as a limiting case of the field equations of the
metric-affine gauge theory of gravity. These two approaches are not
identical in general and coincide only in case when $\Lambda^{\alpha\beta}$
is equal to zero as a consequence of the field equations.

For the variational procedure it is efficiently to use the following general
relations which can be obtained for the arbitrary 2-forms $\Phi ^{\beta
}\!_{\alpha }$, $\Phi _{\alpha }$ and the arbitrary 3-form $\Psi ^{\alpha
\beta }$ with the help of the Cartan structure equations (\ref{eq:304}) and
the structure equation for the nonmetricity 1-form ${\cal Q}_{\alpha \beta }$
(\ref{eq:Q}),
\begin{eqnarray}
\delta {\cal R}^{\alpha }\!_{\beta }\wedge \Phi ^{\beta }\!_{\alpha
}=\rmd (\delta \Gamma ^{\alpha }\!_{\beta }\wedge \Phi ^{\beta }\!_{\alpha
})+\delta \Gamma ^{\alpha }\!_{\beta }\wedge {\rm D}\Phi ^{\beta }\!_{\alpha }
\;,\label{eq:varr} \\
\delta {\cal T}^{\alpha }\wedge \Phi _{\alpha }=\rmd (\delta \theta ^{\alpha
}\wedge \Phi _{\alpha })+\delta \theta ^{\alpha }\wedge {\rm D}\Phi _{\alpha
}+\delta \Gamma ^{\alpha }\!_{\beta }\wedge \theta ^{\beta }\wedge \Phi
_{\alpha }\;,  \label{eq:vart} \\
\delta {\cal Q}_{\alpha \beta }\wedge \Psi ^{\alpha \beta }=\rmd (-\delta
g_{\alpha \beta }\Psi ^{\alpha \beta })+\delta \Gamma ^{\alpha }\!_{\beta
}\wedge 2\Psi _{(\alpha }\!^{\beta )}+\delta g_{\alpha \beta }{\rm D}
\Psi ^{\alpha\beta }\;.  \label{eq:varq}
\end{eqnarray}

The subsequent derivation of the variations of the Lagrangian density 4-form
(\ref{eq:tlag}) is based on the master formula (\ref{eq:master}) derived in
Lemma proved in \cite{bfkl} (see Appendix B). This master formula gives
the rule how to compute the commutator of the variation operator $\delta$
and the Hodge star operator $*$. The result of the variation of every term
of the gravitational field Lagrangian density 4-form (\ref{eq:tlag}) one can
find in Appendix B.

The variation of the total Lagrangian density 4-form (\ref{eq:tot}) with
respect to the Lagrange multiplier 3-form $\Lambda^{\alpha \beta}$ yields
according to (\ref{eq:Lamab}) the Weyl's condition (\ref{eq:Q}) on the
nonmetricity 1-form $Q_{\alpha\beta}$.

The variation of (\ref{eq:tot}) with respect to $\Gamma^\alpha\!_\beta$ can
be obtained by combining all terms in (\ref{eq:lin})--(\ref{eq:Lamab})
proportional to the variation of $\Gamma^\alpha\!_\beta$ and taking into
account the same variation of the fluid Lagrangian density 4-form
(\ref{eq:20}), which is the dilaton-spin momentum 3-form (\ref{eq:d38}).
This variation yields the field $\Gamma$-equation,
\begin{eqnarray}
\fl \frac{1}{4}\lambda \delta_\alpha^\beta \rmd *\! \rmd {\cal Q} -
\frac{1}{8}{\cal Q}\wedge \eta_\alpha\!^\beta + \frac{1}{2}{\cal T}_\gamma
\wedge \eta_\alpha\!^{\beta\gamma} + 2\varrho_1 \theta^\beta \wedge
*{\cal T}_\alpha  \nonumber \\
+ 2\varrho_2 \theta^\beta \wedge \theta_\gamma \wedge *({\cal T}^\gamma
\wedge \theta_\alpha ) + 2\varrho_3 \theta^\beta\wedge \theta_\alpha \wedge
*({\cal T}^\gamma \wedge \theta_\gamma ) \nonumber \\
+ 4\xi \delta_\alpha^\beta *\!{\cal Q} +\zeta \Bigl (\theta^\beta\wedge
*({\cal Q}\wedge \theta_\alpha ) + 2\delta_\alpha^\beta \theta^\gamma \wedge
*{\cal T}_\gamma \Bigr ) - \frac{1}{f_0}\Lambda_\alpha\!^\beta \nonumber \\
\lo = \frac{1}{4f_0}n \left (S^\beta\!_\alpha + \frac{1}{4}J\delta_\alpha^\beta
\right ) u\;, \label{eq:vargg}
\end{eqnarray}
the condition (\ref{eq:Q}) being taken into account after the variational
procedure has been performed.

The variation of (\ref{eq:tot}) with respect to the basis 1-form
$\theta^{\sigma}$ can be obtained in the similar way that gives the second
field equation ($\theta$-equation),
\begin{eqnarray}
\fl \frac{1}{2}{\cal R}^{\alpha }\!_{\beta } \wedge \eta_{\alpha }\!^{\beta }
\!_{\sigma } - \Lambda \eta_\sigma
 + \varrho_1 \Bigl (2 {\rm D}*\!{\cal T}_\sigma + {\cal T}^\alpha \wedge
*({\cal T}_\alpha \wedge \theta_\sigma ) + * (*{\cal T}_\alpha \wedge
\theta_\sigma )\wedge * {\cal T}^\alpha \Bigr )  \nonumber \\
 + \; \varrho_2 \biggl ( 2{\rm D}\Bigl (\theta_{\alpha } \wedge
*({\cal T}^{\alpha }\wedge \theta_{\sigma })\Bigr )
- * ({\cal T}^\beta \wedge \theta_\alpha \wedge \theta_\sigma )
({\cal T}^\alpha \wedge \theta_\beta ) \nonumber \\
 + \; 2{\cal T}^\alpha \wedge *(\theta_\alpha \wedge {\cal T}_\sigma )
- *\Bigl (* ({\cal T}^\beta \wedge \theta_\alpha )\wedge \theta_\sigma \Bigr )
\wedge * ({\cal T}^\alpha \wedge \theta_\beta ) \biggr ) \nonumber \\
 + \; \varrho_3 \biggl ( 2{\rm D}\Bigl (\theta_\sigma \wedge
* ({\cal T}^\alpha \wedge \theta_\alpha )\Bigr ) + 2{\cal T}_\sigma \wedge
* ({\cal T}^\alpha \wedge \theta_\alpha )  \nonumber \\
- *\!({\cal T}^\beta \wedge \theta_\beta \wedge \theta_\sigma )
{\cal T}^\alpha \wedge \theta_\alpha - *\Bigl (*({\cal T}^\beta \wedge
\theta_\beta ) \wedge \theta_\sigma \Bigr ) \wedge * ({\cal T}^\alpha \wedge
\theta_\alpha )\biggr ) \nonumber \\
 + \; \lambda \left (\frac{1}{4}{\cal R}^{\beta }\!_{\beta } \wedge
* ({\cal R}^{\alpha } \!_{\alpha }\wedge \theta_\sigma ) + \frac{1}{4}
*(*{\cal R}^{\alpha } \!_{\alpha }\wedge \theta_\sigma ) \wedge
*{\cal R}^{\beta }\!_{\beta } \right )  \nonumber \\
+ \;\zeta \Bigl ({\rm D}*\!({\cal Q} \wedge \theta_\sigma ) + {\cal Q} \wedge
\theta^\alpha \wedge * ({\cal T}_\alpha \wedge \theta_\sigma )  \nonumber \\
 - \; {\cal Q} \wedge *{\cal T}_\sigma + * (*{\cal T}_{\alpha } \wedge
\theta_{\sigma }) \wedge * ({\cal Q}\wedge \theta^\alpha )\Bigr ) \nonumber \\
 + \;\xi \Bigl (- {\cal Q} \wedge *({\cal Q} \wedge \theta_{\sigma }) -
* (* {\cal Q} \wedge \theta_{\sigma }) *\!{\cal Q} \Bigr )
= - \frac{1}{2f_0}\Sigma_{\sigma}\;.  \label{eq:ein}
\end{eqnarray}
Here $\Sigma_{\sigma}$ is the fluid canonical energy-momentum 3-form
(\ref{eq:d37}). In (\ref{eq:ein}) the condition (\ref{eq:Q}) is used after
the variational procedure has been performed.

      The result of the variation of the total Lagrangian density 4-form
(\ref{eq:tot}) with respect to the metric components $g_{\alpha\beta}$
($g$-equation) is not independent and is a consequence of the field $\Gamma$-
and $\theta$-equations. For the metric-affine theory of gravitation it was
pointed out in \cite{He:pr}. In the Weyl--Cartan theory of gravitation
it can be justified as follows. In this theory, as the consequence of the
scale invariance, the metric of the tangent space can be chosen in the
form \cite{Ut}, $g_{ab} = \sigma (x) g^M_{ab}$, where $g^M_{ab}$ is the metric tensor
of the Minkowski space and $\sigma (x)$ is an arbitrary function to be
varied when the $g$-equation is derived. Therefore the $g$-equation appears
only in the trace form. But the total Lagrangian density 4-form
(\ref{eq:tot}) also obeys the diffeomorphism invariance and therefore
the Noether identity analogous to the identity (\ref{eq:dzak1}) is valid,
from which the trace of the $g$-equation can be derived via the $\Gamma$-
and $\theta$-equations. The quantity $\vec{e}_{\sigma}\rfloor {\cal Q}$ in
(\ref{eq:dzak1}) does not vanish identically in general, otherwise we should
have a Riemann--Cartan spacetime, in which case we could choose
$\sigma (x) = {\rm const} = 1$ and the g-equation would not appear.

\section{\bf The analysis of the field $\Gamma$-equation}
\setcounter{equation}{0}

Let us give the detailed analysis of the $\Gamma$-equation (\ref{eq:vargg}).
The antisymmetric part of this equation determines the torsion 2-form
${\cal T}^\alpha$. The symmetric part determines the Lagrange multiplier
3-form $\Lambda_{\alpha}\!^{\beta}$ and the Weyl 1-form ${\cal Q}$.

After antisymmetrization the equation (\ref{eq:vargg}) gives the following
equation for the torsion 2-form,
\begin{eqnarray}
\fl -\frac{1}{2}{\cal T}^\gamma \wedge \eta_{\alpha\beta\gamma} + \frac{1}{8}
{\cal Q}\wedge \eta_{\alpha\beta} + 2\varrho_1 \theta_{[\alpha}\wedge *{\cal T}
_{\beta ]}  \nonumber \\
+ 2\varrho_2 \theta_{[\alpha} \wedge \theta_{\mid\gamma
\mid }\wedge *({\cal T}^{\mid\gamma\mid}\wedge \theta_{\beta ]})
+ 2 \varrho_3 \theta_\alpha \wedge \theta_\beta \wedge *({\cal T}^\gamma
\wedge \theta_\gamma ) \nonumber\\
+ \zeta \theta_{[\alpha}\wedge *({\cal Q}\wedge \theta_{\beta ]}) =
\frac{1}{2}\mbox{\ae} n S_{\alpha\beta } u\;, \qquad \mbox{\ae} =
\frac{1}{2f_0} \;. \label{eq:Tor}
\end{eqnarray}

The torsion 2-form can be decomposed into the irreducible pieces (the
traceless 2-form $\stackrel{(1)}{{\cal T}}\!^{\alpha}$, the trace 2-form
$\stackrel{(2)}{{\cal T}}\!^{\alpha}$ and the pseudotrace 2-form $\stackrel
{(3)}{{\cal T}}\!^{\alpha}$) \cite{He:pr}, \cite{Ob-Hel},
\begin{equation}
{\cal T}^{\alpha} = \stackrel{(1)}{{\cal T}}\!^{\alpha} + \stackrel{(2)}
{{\cal T}}\!^{\alpha} + \stackrel{(3)}{{\cal T}}\!^{a}\; .  \label{eq:razl}
\end{equation}
Here the torsion trace 2-form and the torsion pseudotrace 2-form of the
pseudo-Rieman\-ni\-an 4-manifold are determined by the expressions,
respectively,
\begin{eqnarray}
\stackrel{(2)}{{\cal T}}\!^{\alpha} = \frac{1}{3}{\cal T}\wedge\theta^{\alpha} \;,
\qquad &{\cal T} = * (\theta_\alpha \wedge *{\cal T}^{\alpha} ) =
-(\vec e_{\alpha} \rfloor {\cal T}^{\alpha})\;, \label{eq:sled} \\
\stackrel{(3)}{{\cal T}}\!^{\alpha} = \frac{1}{3}*({\cal P}\wedge
\theta^{\alpha})\;, \qquad
&{\cal P} = * (\theta_\alpha \wedge {\cal T}^{\alpha}) = \vec e_{\alpha}
\rfloor *{\cal T}^{\alpha} \;,  \label{eq:psled}
\end{eqnarray}
where the torsion trace 1-form ${\cal T}$ and the torsion pseudotrace 1-form
${\cal P}$ are introduced.

The irreducible pieces of torsion satisfy to the conditions \cite{He:pr},
\begin{eqnarray}
\stackrel{(1)}{{\cal T}}\!^{\alpha }\wedge {\theta }_{\alpha }=0\;,\qquad
&\stackrel{(2)}{{\cal T}}\!^{\alpha }\wedge {\theta }_{\alpha }=0\;,
\label{eq:sv1} \\
\vec{e}_{\alpha }\rfloor \stackrel{(1)}{{\cal T}}\!^{\alpha }=0\;,\qquad
&\vec{e}_{\alpha }\rfloor \stackrel{(3)}{{\cal T}}\!^{\alpha }=0\;.
\label{eq:sv2}
\end{eqnarray}

Using the computational rules (\ref{eq:eta1})--(\ref{eq:svp}) and
(\ref{eq:rul2}) let us derive two efficient identities,
\begin{eqnarray}
\fl {\cal T}^{\gamma }\wedge \eta_{\alpha\beta\gamma} = {\cal T}^{\gamma }
\wedge (\vec {e}_\gamma\rfloor \eta_{\alpha\beta}) = \vec {e}_\gamma\rfloor
({\cal T}^{\gamma }\wedge \eta_{\alpha\beta}) - (\vec {e}_\gamma\rfloor
{\cal T}^{\gamma }) \wedge \eta_{\alpha\beta} \nonumber \\
\lo = \vec {e}_\gamma\rfloor \Bigl ({\cal T}^\gamma \wedge *(\theta_\alpha
\wedge \theta_\beta )\Bigr ) + {\cal T}\wedge \eta_{\alpha\beta} \nonumber \\
\lo = \Bigl (\vec {e}_\gamma\rfloor (\theta_\alpha\wedge\theta_\beta )\Bigr )
\wedge *{\cal T}^\gamma + \theta_\alpha \wedge \theta_\beta \wedge (\vec {e}
_\gamma\rfloor *{\cal T}^\gamma ) +{\cal T}\wedge \eta_{\alpha\beta}\nonumber\\
\lo = - \theta_{[\alpha}\wedge *{\cal T}_{\beta ]} + \theta^{\alpha }\wedge
\theta_\beta \wedge {\cal P} + {\cal T}\wedge \eta_{\alpha\beta}\;,
\label{eq:id1} \\
\fl \theta_\gamma \wedge *({\cal T}^\gamma \wedge
\theta_\alpha ) = *\Bigl (\vec {e}_\gamma\rfloor ({\cal T}^\gamma \theta
_\alpha )\Bigr ) = * (-{\cal T}\wedge \theta_\alpha + {\cal T}_\alpha ) =
*{\cal T}_\alpha - 3\,*\!\stackrel{(2)}{{\cal T}}\!_{\alpha }\;.\label{eq:id2}
\end{eqnarray}

    Using the identities (\ref{eq:id1}), (\ref{eq:id2}), one can represent
the field equation (\ref{eq:Tor}) as follows,
\begin{eqnarray}
\fl (1 + 2\varrho_1 + 2\rho_2 )\theta_{[\alpha}\wedge *{\cal T}_{\beta ]}
+\left(-\frac{1}{2} + 2\varrho_3 \right ) \theta_\alpha \wedge \theta_\beta
\wedge {\cal P} - 6\varrho_2 \theta_{[\alpha}\wedge *\!\stackrel{(2)}{{\cal T}}\!
_{\beta ]} \nonumber \\
- \frac{1}{2}{\cal T} \wedge
\eta_{\alpha\beta} + \frac{1}{8}{\cal Q}\wedge \eta_{\alpha\beta} +
\zeta \theta_{[\alpha }\wedge *({\cal Q}\wedge \theta_{\beta ]}) =
\frac{1}{2}\mbox{\ae} nS_{\alpha\beta } u \; .  \label{eq:Tor1}
\end{eqnarray}

    Multiplying the equation (\ref{eq:Tor1}) by $\theta^\beta$ from the right
externally, using the com\-pu\-ta\-tion rules (\ref{eq:com1})--(\ref{eq:rul3})
and then the Hodge star operation, one gets in consequence of the Frenkel
condition the relation between the torsion trace 1-form ${\cal T}$
(\ref{eq:sled}) and the Weyl 1-form ${\cal Q}$,
\begin{equation}
{\cal T} = \frac{3(\frac{1}{4} + \zeta)}{2(1 - \varrho_1 + 2\varrho_2 )}
{\cal Q}\; .  \label{eq:con}
\end{equation}

As a consequence of (\ref{eq:con}) and the relation
(\ref{eq:teta4}) it can be proved the equality for the trace 2-form,
\begin{equation}
\fl (1 +2\varrho_1 -4\varrho_2 )\theta_{[\alpha}\wedge *\!\stackrel{(2)}{{\cal
T}}\!_{\beta ]} - \frac{1}{2}{\cal T}\wedge \eta_{\alpha\beta} + \frac{1}{8}
{\cal Q}\wedge \eta_{\alpha\beta} + \zeta\theta_{[\alpha }
\wedge *({\cal Q}\wedge \theta_{\beta ]}) = 0\;. \label{eq:Tor2}
\end{equation}

    Then as a consequence of (\ref{eq:razl}), (\ref{eq:psled}),
(\ref{eq:Tor2}) and the conditions (\ref{eq:sv1}) the field equation
(\ref{eq:Tor1}) is transformed as follows,
\[
\fl (1 +2\varrho_1 +2\varrho_2 )\,\theta_{[\alpha}\wedge *\!\stackrel{(1)}
{{\cal T}}\!_{\beta ]} - \frac{1}{6}(1 -4\varrho_1 -4\varrho_2 -12\varrho_3)
\theta_\alpha\wedge \theta_\beta \wedge {\cal P} = \frac{1}{2}\mbox{\ae}
nS_{\alpha\beta } u \;.
\]

    Contracting this equation with $g^{\beta\gamma}\vec e_\gamma$, we get
with the help of the Leibnitz rule (\ref{eq:Lrule}) the equation,
\begin{equation}
\fl (1 +2\varrho_1 +2\varrho_2 ) *\!\stackrel{(1)}{{\cal T}}\!_{\alpha} -
\frac{2}{3}(1 -4\varrho_1 -4\varrho_2 -12\varrho_3 )\theta_\alpha \wedge
{\cal P} =\mbox{\ae} nS_{\alpha\beta } u_\gamma \eta^{\beta\gamma}\,.
\label{eq:Tor3}
\end{equation}
By contracting this equation with $g^{\alpha\beta}\vec e_\beta$ we get
the equation
\begin{equation}
(1 -4\varrho_1 -4\varrho_2 -12\varrho_3 ) {\cal P} = \mbox{\ae} n \sigma\;,
\label{eq:P}
\end{equation}
which represents the torsion pseudotrace 1-form ${\cal P}$ via the
Pauli--Lyubanski spin 1-form $\sigma$ of a fluid particle,
\begin{equation}
\sigma = - \frac{1}{2}S^{\alpha\beta} u^\gamma \eta_{\alpha\beta\gamma} =
\frac{1}{2}S^{\alpha\beta} u^\gamma \eta_{\lambda\alpha\beta\gamma}
\theta^\lambda \;.  \label{eq:PL}
\end{equation}

    As a consequence of (\ref{eq:P}) the field equation (\ref{eq:Tor3})
yields the equation for the traceless piece of the torsion 2-form,
\begin{eqnarray}
\fl (1 +2\varrho_1 + 2\varrho_2 )\stackrel{(1)}{{\cal T}}\!_{\alpha} =
\mbox{\ae}n \left ( S_{\alpha}\!_{\beta}u_\gamma \theta^\beta \wedge
\theta^\gamma + \frac{2}{3}\sigma^\beta \eta_{\beta\alpha} \right ) =
- \frac{2}{3}\mbox{\ae} n S_{\beta (\alpha}u_{\gamma )}
\theta^\beta \wedge \theta^\gamma \;. \label{eq:T1}
\end{eqnarray}

    Now let us calculate the symmetric part of the $\Gamma$-equation
(\ref{eq:vargg}). Because of (\ref{eq:svlam}) the result can be represented
as follows,
\begin{eqnarray}
\fl \mbox{\ae}  \Lambda_{\alpha\beta} =
\varrho_1 \theta_{(\alpha }\wedge *{\cal T} _{\beta )} +
\varrho_2 \theta_{(\alpha }\wedge \theta_{\mid\gamma\mid} \wedge
*({\cal T}^{\mid\gamma\mid} \wedge \theta_{\beta )})
+ \frac{1}{8} \lambda g_{\alpha\beta} \rmd *\!\rmd {\cal Q}
+ 2\xi g_{\alpha\beta} *\!{\cal Q} \nonumber \\
+ \zeta \left (\frac{1}{2} \theta_{(\alpha }\wedge *({\cal Q} \wedge
\theta_{\beta )}) + g_{\alpha\beta} \theta_{\gamma}\wedge *{\cal T}_\gamma
\right ) - \frac{1}{32}\mbox{\ae} n g_{\alpha\beta} Ju \;.  \label{eq:sym}
\end{eqnarray}

    By contracting the equation (\ref{eq:sym}) on the indices $\alpha $ and
$\beta $ and after substituting (\ref{eq:con}) in the result, one finds the
equation of the Proca type for the Weyl 1-form,
\begin{equation}
\fl *\rmd * \!\rmd {\cal Q} + m^{2} {\cal Q} = \frac{{\mbox \ae}}{2\lambda}
 n J *\!u \;, \qquad
m^{2} = 16\frac{\xi }{\lambda } + \frac{3(\varrho _1 - 2\varrho_2 + 8\zeta
(1 + 2\zeta ))}{4\lambda (1 -\varrho_1 +2\varrho_2 )}\;.  \label{eq:dQ}
\end{equation}
The equation (\ref{eq:dQ}) shows that the dilaton field ${\cal Q}$, in
contrast to Maxwell field, possesses the non-zero rest mass and
demonstrates a short-range nature, as it was pointed out by Utiyama
\cite{Ut} (see also \cite{Sola}). In the component form the Proca type
equation for Weyl vector was used in \cite{Ut,Arg} and in the exterior
form language in \cite{Ob-Hel}.

    By virtue of $\rmd (nu) = 0$ and $\dot{J} = 0$ (see section 2) the
equation (\ref{eq:dQ}) has the Lorentz condition as a consequence,
\begin{equation}
 \rmd\, *\! {\cal Q} = 0  \;, \qquad \stackrel{R}{\nabla}_\alpha Q^\alpha = 0
\;, \label{eq:d*Q}
\end{equation}
where $\stackrel{R}{\nabla}_\alpha$ is the covariant derivative with respect
to the Riemann connection (see Appendix C). Here the latter relation is the
component representation of the former one.

    If we use (\ref{eq:con}) and (\ref{eq:dQ}), then the equation
(\ref{eq:sym}) takes the form,
\begin{eqnarray}
\fl \mbox{\ae} \Lambda_{\alpha\beta} =
\varrho_1 \theta_{(\alpha }\wedge *{\cal T}_{\beta )}
+ \varrho_2 \theta_{(\alpha }\wedge \theta_{\mid\gamma\mid} \wedge
*({\cal T}^{\mid\gamma\mid} \wedge \theta_{\beta )})  \nonumber \\
+ \frac{4\zeta (1 -\varrho_1 + 2\varrho_2)}{3(1 + 4\zeta )}\theta_{(\alpha }
\wedge *({\cal T} \wedge \theta_{\beta )}) -\frac{4\zeta +\varrho_1
- 2\varrho_2}{4(1 + 4\zeta )}g_{\alpha\beta} *\!{\cal T}\;. \label{eq:multy}
\end{eqnarray}
This equation determines the Lagrange multiplier 3-form
$\Lambda_{\alpha\beta}$. It is very important that $\Lambda_{\alpha\beta}$
is in general not equal to zero.

    The equations (\ref{eq:con}), (\ref{eq:P}), (\ref{eq:T1}) and
(\ref{eq:multy}) solve the problem of the evaluation the torsion 2-form and
the Lagrange multiplier 3-form. With the help of the algebraic field
equations (\ref{eq:P}) and (\ref{eq:T1}) the traceless and pseudotrace
pieces of the torsion 2-form are determined via the spin tensor and the flow
3-form $u$ of the perfect dilaton-spin fluid in general case, when the
conditions $1 +2\varrho_1 +2\varrho_2 \not = 0$ and $1 -4\varrho_1 -4\varrho_2
- 12\varrho_3 \not = 0$ are valid. With the help of the equation (\ref{eq:con})
one can determine the torsion trace 2-form via the dilaton field ${\cal Q}$,
for which the differential field equation (\ref{eq:dQ}) is valid. Therefore
the torsion trace 2-form can propagate in the theory under consideration.

\section{\bf The field $\theta$-equation in homogeneous and isotropic
          cosmology}
\setcounter{equation}{0}

Now we consider the homogeneous and isotropic Universe filled with the
perfect dilaton-spin fluid, which realizes the model of the dark matter with
$J \not = 0$ in contrast to the ordinary baryonic matter with $J = 0$. The
metric of this cosmological model is the Robertson--Walker (RW) metric with
scale factor $a(t)$,
\begin{equation}
\rmd s^{2} = \frac{a^{2} (t)}{1 - kr^{2}}\rmd r^{2} + a^{2} (t)r^{2}
\Bigl (\rmd \theta^2 + (\sin{\theta})^2 \rmd \phi^2 \Bigr ) - \rmd t^{2}\;,
\label{eq:RW}
\end{equation}
the comoving frame of reference being chosen,
\begin{equation}
u^1 = u^2 = u^3 = 0 \;, \qquad u^4 = 1\;. \label{eq:ref}
\end{equation}

    As it was shown in \cite{Tsam,Min1}, in the spacetime with the RW
metric (\ref{eq:RW}) the only nonvanishing components of the torsion are
$T^{1}\!_{41} = T^{2}\!_{42} = T^{3}\!_{43}$ and $T_{ijk}$ for
$i = 1,\;2,\;3$. In this case from (\ref{eq:sled}) we get that the only
nonvanishing component of the trace 1-form is $T_{4} = T_{4} (t)$
($T_i = 0$ for $i = 1,\;2,\;3$). From (\ref{eq:psled}) we also find,
${\cal P} = 3T_{[123]}\eta^{1234}\theta_4$. But the field equation
(\ref{eq:P}) yields, ${\cal P}^4 \sim \sigma^4 = 0$, as a consequence of
(\ref{eq:PL}) and (\ref{eq:ref}). Therefore the pseudotrace piece of the
torsion 2-form vanishes. It is easy to calculate with the help of
(\ref{eq:razl}), (\ref{eq:sled}) that the traceless piece also vanishes.
Therefore for the RW metric (\ref{eq:RW}) we have,
\begin{equation}
\stackrel{(1)}{{\cal T}}\!_{\alpha} = 0\;, \qquad
\stackrel{(3)}{{\cal T}}\!_{\alpha} = 0\;, \label{eq:T130}
\end{equation}
and the torsion 2-form consists only from the trace piece that in the
component repre\-sen\-ta\-ti\-on reads,
\begin{equation}
T_{\lambda\alpha\beta} = - \frac{2}{3} g_{\lambda [\alpha} T_{\beta ]} \;.
\label{eq:T}
\end{equation}

     As a consequence of (\ref{eq:P})--(\ref{eq:T1}) and the identity,
\begin{equation}
u_{\lambda}S_{\alpha\beta} \equiv u_{[\lambda}S_{\alpha\beta ]} + \frac{2}{3}
(u_{(\lambda}S_{\alpha )\beta} - u_{(\lambda}S_{\beta )\alpha} )\;,
\label{eq:eq}
\end{equation}
we have to conclude that the condition $S_{\alpha\beta} = 0$ is valid for the
spin tensor of the matter source in the cosmological model considered. It
can be understood in the sense that the mean value of the spin tensor is
equal to zero under statistical averaging over all directions in the
homogeneous and isotropic Universe. As a consequence of this fact in this
section and in the sequential sections we shall simplify the equations of
the theory by using the conditions (\ref{eq:T130}) and $S_{\alpha\beta} =
0$. In case $S_{\alpha\beta} = 0$ dilaton-spin fluid becomes dilaton fluid.

      Let us now decompose the field $\theta$-equation (\ref{eq:ein}) into
Riemannian and non-Riemannian parts using the formulae
(\ref{eq:Grusl})--(\ref{eq:red3}) and then transform the result to the
component form. In Appendix C one can find the decomposition of the each
part of the field equation (\ref{eq:ein}). After gathering all expressions
(\ref{eq:f1})--(\ref{eq:f5}) together and substituting the relation
(\ref{eq:con}) we receive the following results.

    The terms with the derivatives of the dilaton field, like
$\stackrel{R}{\nabla}_{\alpha} Q^{\alpha}$ and
$\stackrel{R}{\nabla}_{\sigma} Q^{\alpha}$, and the same derivatives of the
torsion trace 1-form in a remarkable manner mutually compensate each other
and vanish as a consequence of (\ref{eq:con}),
\[
\fl \frac{2}{3} (1 -\varrho_1 +2\varrho_2 )\left (\eta_\sigma\stackrel{R}
{\nabla}_{\rho}T^\rho - \eta_\rho \stackrel{R}{\nabla}_{\sigma}
T^\rho \right ) - \left (\frac{1}{4} + \zeta \right ) \left (\eta_\sigma
\stackrel{R}{\nabla}_\rho Q^\rho - \eta_\rho \stackrel{R}{\nabla}_\sigma
Q^\rho \right ) = 0 \;.
\]

    The terms with $\rmd {\cal Q}$ also vanish, as the equality
$\rmd {\cal Q} = 0$ is valid identically for the RW metric (\ref{eq:RW})
that can be easy verified in the holonomic basis, when
$\theta^\alpha = \rmd x^\alpha$,
\begin{equation}
\rmd {\cal Q} = \partial_\beta Q_\alpha \rmd x^\beta \wedge \rmd x^\alpha =
\partial_{4} Q_{4} \rmd x^{4} \wedge \rmd x^{4} = 0 \;.  \label{eq:dQ0}
\end{equation}
This follows from the fact that for this metric one has $Q_{4} = Q_{4}(t)$,
$Q_i = 0$ ($i = 1,\,2,\,3$) as a consequence of (\ref{eq:con}) and the values
of the trace torsion for the metric (\ref{eq:RW}).

      The remainder terms of the equation (\ref{eq:ein}) after some algebra
can be represented as follows,
\begin{eqnarray}
\left (\stackrel{R}{R}\!^{\rho}\!_{\sigma} -\frac{1}{2}\delta^\rho_\sigma
\stackrel{R}{R}\right ) \eta_\rho + \Lambda\eta_\sigma +
\alpha (2Q_\sigma Q^\rho \eta_\rho - Q_\rho Q^\rho \eta_\sigma ) =
\mbox{\ae}\Sigma_\sigma\;, \label{eq:res}\\
\alpha =\frac{3\left (\frac{1}{4} +\zeta \right )^2}
{4(1 - \varrho_1 + 2\varrho_2 )} + \xi - \frac{3}{64}\;. \label{eq:alf}
\end{eqnarray}

      Then we shall derive the Weyl 1-form ${\cal Q}$ algebraically as a
consequence of (\ref{eq:dQ0}) from the equation (\ref{eq:dQ}) via the right
side of this equation,
\begin{equation}
Q^\alpha = \frac{\mbox{\ae}}{2\lambda m^2}nJu^\alpha \;,  \label{eq:qa}
\end{equation}
which is in accordance with the conditions (\ref{eq:ref}) for the
comoving system of reference.

      After substituting (\ref{eq:qa}) and (\ref{eq:d37}) to the equation
(\ref{eq:res}), the condition $S_{\alpha\beta} = 0$ being used, we can
represent the field equation (\ref{eq:ein}) as an Einstein-like equation,
\begin{equation}
\stackrel{R}{R}_{\sigma\rho} - \frac{1}{2}g_{\sigma\rho}\stackrel{R}{R} =
 \mbox{\ae} \Bigl ((\varepsilon_{\rm {e}} +p_{\rm {e}} ) u_\sigma u_\rho +
p_{\rm{e}}g_{\sigma\rho}\Bigr )\;,  \label{eq:ein1}
\end{equation}
where $\stackrel{R}{R}_{\sigma\rho}$, $\stackrel{R}{R}$ are a Ricci tensor
and a curvature scalar of a Riemann space, respectively,
$\varepsilon_{\rm e}$ and $p_{\rm e}$ are an energy density and a pressure
of an effective perfect fluid:
\begin{equation}
\varepsilon_{\rm{e}} = \varepsilon + \varepsilon_{\rm{v}} - \alpha \mbox{\ae}
\left (\frac{nJ}{2 \lambda m^2}\right )^2\;, \qquad
p_{\rm{e}} = p + p_{\rm{v}} - \alpha \mbox{\ae} \left (\frac{nJ}{2 \lambda m^2}\right )^2
\;, \label{eq:eff}
\end{equation}
and $\varepsilon_{\rm v} = \Lambda /\mbox{\ae}$ and $p_{\rm v} =
-\Lambda /\mbox{\ae}$ are an energy density and a pressure of a vacuum with
the equation of state, $\varepsilon_{\rm v} = -p_{\rm v} > 0 $.

\section{\bf Evolution of the Universe with dilaton matter}
\setcounter{equation}{0}

The field equation (\ref{eq:ein1}) yields the modified
Friedmann--Lema\^{\i}tre (FL) equation,
\begin{equation}
\left (\frac{\dot a}{a}\right )^2 + \frac{k}{a^2} = \frac{\mbox{\ae}}{3}\left
(\varepsilon + \varepsilon_{\rm{v}} - \alpha\mbox{\ae} \left (\frac{Jn}{2
\lambda m^2}\right )^2 \right )\;.   \label{eq:frid}
\end{equation}

    The integration of the continuity equation $\rmd (nu)=0$ ($\rmd$ -- the
operator of exterior differentiation) for RW metric (\ref{eq:RW}) yields the
matter conservation law $na^3 = N = {\rm const}$. As an equation of state of
the dilaton fluid we choose the equation of state
\begin{equation}
p = \gamma \varepsilon \;, \qquad 0 \le \gamma < 1\;. \label{eq:eqst}
\end{equation}
Then the integration of the energy conservation law
(\ref{eq:cons}) for RW metric (\ref{eq:RW})  yields the condition,
\begin{equation}
\varepsilon\, a^{3(1 +\gamma )} = {\cal E}_\gamma = {\rm const} \;, \qquad
{\cal E}_\gamma >0 \;. \label{eq:E}
\end{equation}

    By virtue of these relations the modified FL equation (\ref{eq:frid})
takes the form,
\begin{equation}
\fl \left (\frac{\dot a}{a} \right )^2  + \frac{k}{a^2} = \frac{\mbox{\ae}}
{3a^6} \left ( \varepsilon_{\rm{v}} a^6 + {\cal E}_\gamma a^{3(1-\gamma)}
- {\cal E}\right )\;, \qquad {\cal E} = \alpha\mbox{\ae}\left (\frac{JN}
{2\lambda m^2}\right )^2  \;. \label{eq:fridd}
\end{equation}
From this equation one can see that the influence of dilaton matter is most
essential at the early stage of the Universe evolution (when $a<<1$), and
that the vacuum energy contribution dominates at the last stage $a>>1$,
when the size of the Universe will exceed some specific magnitude defined
by the parameters of the dilaton fluid and dilaton field.

     Now we put $k = 0$ in (\ref{eq:fridd}) in accordance with the modern
observational evidence \cite{nabl,Os-Perl,St}, from which one should conclude
that the Universe is spatially flat in cosmological scale. In this case we can
conclude from the equation (\ref{eq:fridd}) that extremum points of the scale
factor ($\dot a = 0$) correspond to zero points of the equation,
\begin{equation}
a^6  + \frac{{\cal E}_\gamma}{\varepsilon_{\rm{v}}} \left (a^{3(1-\gamma)} -
\frac{{\cal E}}{{\cal E}_\gamma} \right ) = 0 \;. \label{eq:zer}
\end{equation}

    We are finding zero points corresponding to the positive values of the
scale factor, $a > 0$. It is easy to see that, if the condition,
\begin{equation}
0 < {\cal E} << 1\;, \label{eq:usl}
\end{equation}
is valid, then in case $a<<1$ the equation (\ref {eq:zer}) has one such zero
point,
\begin{equation}
a_{0} \approx \left (\frac{{\cal E}}{{\cal E}_\gamma} \right )^{\frac{1}
{3(1-\gamma)}} = \left (\frac{\alpha\mbox{\ae}}{{\cal E}_\gamma}\left (
\frac{J N}{2\lambda m^2}\right )^2 \right )^{\frac{1}{3(1-\gamma)}}\;,
\label{eq:a0}
\end{equation}
and none zero points in case $a>>1$ (as ${\cal E}_\gamma >0$).

   In order to clarify, whether there is a minimum or a maximum in the
extremum point, one can use the other components of the equation
(\ref{eq:ein1}). By virtue of (\ref{eq:E}) and (\ref{eq:a0}) the result reads,
\begin{equation}
\fl\frac{\ddot a}{a} = - \frac{\mbox{\ae} }{6} (\varepsilon_{\rm{e}} +
3p_{\rm{e}}) = \frac {\mbox{\ae} }{3a^6} \left (\varepsilon_{\rm{v}}a^6 -
\frac{1}{2}(1+3\gamma ){\cal E}_\gamma a^{3(1-\gamma)} + 2{\cal E} \right )\;.
\label{eq:add}
\end{equation}

    At the extremum point (\ref{eq:a0}) (in case $a<<1$) one gets
\[
\left (\frac{\ddot a}{a}\right)_{a=a_0} = \frac{\mbox{\ae}}{3}\left
(\varepsilon_{\rm v} + \frac{3(1-\gamma)}{2}\frac{{\cal E}}{a_0^{6}} \right )
> 0 \;,
\]
Therefore the value $a_{0}$ corresponds to the minimum point of the scale
factor $a(t)$.

    We conclude that under the some conditions on the parameters
of the Lagrangian density 4-form (\ref{eq:tlag}) there exists a nonsingular
solution of the homogeneous and isotropic cosmological model of the Universe
to which the minimum value of the scale factor $a_{0}$
and the maximum value of the matter density of the Universe correspond,
\begin{equation}
\varepsilon_{\rm{max}} \approx \frac{{\cal E}_\gamma}{a_0^{3(1+\gamma)}} =
\frac{{\cal E}}{a_0^6} = \left (\frac{{\cal E}^2_\gamma}
{{\cal E}^{1+\gamma} }\right )^{\frac{1}{1-\gamma}}\;. \label{eq:max}
\end{equation}
In case $\gamma = 0$ the value (\ref{eq:max}) corresponds to that one which
was established in \cite{bfGrC}, where the influence of the vacuum energy
contribution was neglected.

    In case $a>>1$ we can neglect the last term in (\ref{eq:add}) and get
the equation,
\begin{equation}
\frac{\ddot a}{a} = \frac{\mbox{\ae} }{3} \left (\varepsilon_{\rm v} -
\frac{1}{2}(1+3\gamma ) \varepsilon (t) \right )\;, \label{eq:add1}
\end{equation}
where $\varepsilon (t)$ is the current value of the dilaton fluid energy
density. This equation is valid for the most part of the history of the
Universe.

    Consider now the conditions under which the points of inflection of the
function $a(t)$ can exist. To this end let us equate the right side of
(\ref{eq:add}) to zero ($\ddot a = 0$) and find zero points of the equation,
\begin{equation}
a^6 - \frac{(1+3\gamma ){\cal E}_\gamma}{2\varepsilon_{\rm{v}}}\left (
a^{3(1-\gamma)} - \frac{4}{1+3\gamma}a_{0}^{3(1-\gamma)} \right ) = 0\;.
\label{eq:add0}
\end{equation}
As the last term in (\ref{eq:add0}) is very small, then this equation has
two types of zero points: very small by magnitude with the value,
\begin{equation}
a_1 \approx a_0\left (\frac{4}{1+3\gamma} \right )^{\frac{1}{3(1-\gamma)}} \;,
\label{eq:a1}
\end{equation}
and large by magnitude with the value,
\begin{equation}
a_2 \approx \left (\frac{(1+3\gamma){\cal E}_\gamma}{2\varepsilon_{\rm v}}
\right )^{\frac{1}{3(1+\gamma)}} \; .  \label{eq:a2}
\end{equation}

    As both solutions correspond to the positive values of the scale
factor ($a>0$), then two points of inflection exist. As $\gamma < 1$, the
first point of inflection (\ref{eq:a1}) occurs at once after the minimum
$a_0$ of the scale factor. Then up to the second point of inflection $a_2$
one has $\ddot{a} < 0$ and the expansion with deceleration occurs up to the
end of the Friedmann era. The point of inflection $a_2$ corresponds to the
modern era when the Friedmann expansion with deceleration is replaced
by the expansion with acceleration that means the beginning of the "second
inflation" era.

    By equating to zero the equation (\ref{eq:add1}) we get the correlation
between the vacuum energy density $\varepsilon_{\rm v}$ and the dilaton
matter energy density $\varepsilon$  at the point of inflection $a_2$,
\begin{equation}
\varepsilon = \frac{2\varepsilon_{\rm v}}{1+3\gamma }  \;. \label{eq:usl2}
\end{equation}

    After the second point of inflection $a_2$ the dilaton matter energy
density $\varepsilon$ diminishes and the inequality is valid,
\[
\varepsilon < \frac{2\varepsilon_{\rm v}}{1+3\gamma }  \;,
\]
that corresponds to the condition $\ddot {a} >0$ and therefore to the
expansion with acceleration.

     For the nonrelativistic cold matter ($\gamma = \case23 $) the formula
(\ref{eq:usl2}) yields,
\[
\Omega_{\rm cdm} = \frac{2}{3} \Omega_\Lambda \;,  \qquad
\Omega_{\rm cdm} = \frac{\varepsilon}{\varepsilon_{\rm tot}}\;, \qquad
\Omega_\Lambda = \frac{\varepsilon_{\rm v}}{\varepsilon_{\rm tot}}\;, \qquad
\varepsilon_{\rm tot} = \frac{3H^2}{8\pi G}\; .
\]
that fits to the boundary of the modern observational data \cite{nabl},
\[
\Omega_\Lambda = 0.66 \pm 0.06 \;, \qquad \Omega_{\rm cdm} h_0^2 =
0.17 \pm 0.02\;,
\]
with $H_0 =100\,h_0 =65\;km\;s^{-1}\,Mpc^{-1}$ \cite{Os-Perl}.

  Curiously, that if one takes the generally accepted data of \cite{Os-Perl},
$\Omega_{\rm cdm} = \frac{1}{3}\,, \;\Omega_\Lambda= \frac{2}{3}$, and
substitute to (\ref{eq:usl2}), then one gets the value $\gamma = 1$, that
corresponds to the equation of state of the superrigid matter.
Therefore we can conclude that the theory together with the observational
data gives the approximate range for the viable values of the factor
$\gamma$ in the equation of state of the dilaton matter,
\begin{equation}
\frac{2}{3} \le \gamma \le 1 \;. \label{eq:gamma}
\end{equation}

     It is interesting to investigate the limiting case $\gamma = 1$.  For
this case the equations (\ref{eq:E})--(\ref{eq:zer}) are valid, but the
zero point of the equation (\ref{eq:zer}) in case $\gamma = 1$ is
\begin{equation}
a_{{\rm e}} = \left (\frac{{\cal E} - {\cal E}_1}{\varepsilon_{{\rm v}}}
\right )^{\frac{1}{6}}\;,  \label{eq:extr}
\end{equation}
where ${\cal E}_1 = {\cal E}_{\gamma =1}$ is the integration constant of
the equation (\ref{eq:E}).

  In case $\gamma = 1$ the equation (\ref{eq:add}) at the extremum point
(\ref{eq:extr}) yields
\[
\left (\frac{\ddot a}{a}\right )_{a=a_{{\rm e}}} = \frac {\mbox{\ae}}
{3a_{{\rm e}}^6} \left (\varepsilon_{\rm{v}} a_{{\rm e}}^6 - 2{\cal E}_1 +
2{\cal E} \right ) = 3\Lambda > 0 \;.
\]
Therefore the value $a_{{\rm e}}$ corresponds to the minimum point of
the scale factor, $a_{{\rm e}} = a_{{\rm min}}$.

    In case $\gamma = 1$, $k = 0$ the equation (\ref{eq:fridd}) reads,
\[
\left (\frac{\dot a}{a} \right )^2  = \frac{\mbox{\ae}}{3a^6}
\left (\varepsilon_{\rm{v}}a^6 - {\cal E}_1 + {\cal E} \right ) =
\frac{\Lambda}{3a^6} \left (a^6 - a^6_{{\rm min}} \right )\;.
\]
This equation can be exactly integrated. The solution corresponding to the
initial data $t=0$, $a = a_{{\rm min}}$ reads,
\[
a = a_{{\rm min}} (\cosh{\sqrt{3\Lambda}\,t})^{\frac{1}{3}}\;, \qquad
a_{{\rm min}} = \left (\frac{\alpha {\mbox{\ae}}^2}{\Lambda} \left (\frac{JN}
{2\lambda m^2}\right )^2 - \frac{\mbox{\ae} {\cal E}_1}{\Lambda}
\right )^{\frac{1}{6}}\;.
\]
This solution describes the inflation-like stage of the evolution of the
Universe, which continues until the equation of state of the dilaton matter
will change and will become differ from the equation of state of the
superrigid matter.

\section{Discussion}
\setcounter{equation}{0}

     We come to the conclusion that the matter with dilaton charge can
play the essential role in the whole dynamic of the Universe. The existence
of such matter at the beginning of the Universe leads, if the condition
(\ref{eq:usl}) is fulfilled, to the absence of the initial singularity in
the cosmological solution of the gravitational theory in a Weyl--Cartan
spacetime and to the existence of the maximum value of the matter density of
the Universe, in contrast to the opinion of \cite{Ob-Hel}.  In \cite{Min1}
the nonsingular solution of the cosmological problem was established
in the metric--affine theory of gravitation with quadratic Lagrangians. In
the theory under consideration the nonsingular solution appears not as a
consequence of the quadratic Lagrangians by itself, but because of the
existence of the dilaton matter at the early stage of the Universe.

    The condition (\ref{eq:usl}) means that $\alpha > 0$ and the quantity
$|\lambda m^2 |$ is very large by magnitude. From relation
(\ref{eq:dQ}) it follows that $m^2$ is determined by the quantities $\xi
/\lambda$ and $\zeta^2 /\lambda$ (if $\rho_i \approx 1\,,\; i=1,\,2,\,3\,)$.
Therefore the mass $m$ of a quantum of the dilaton field and at least one of
the coupling constants, $\xi$ or $\zeta$, has to be very large. As
$\alpha > 0$, then we see from (\ref{eq:alf}) that it is $\zeta$. From
(\ref{eq:con}) it follows that the large value of $\zeta$ will intensify to a
great extent the value of a torsion field, the coupling constant of which
with matter is very small in the absence of the dilaton field. In this case
torsion can have any influence on Particle Physics. In other
words, the great dilaton field mass specifies the extent of the influence
of the torsion field on the matter dynamics that finally determines the
minimal radius of the Universe, the initial point of the Friedmann era and
maybe some properties of elementary particles behavior.

    The condition (\ref{eq:usl}) also guarantees the existence of two points
of inflection of the scale factor function $a(t)$, between which the Friedmann
era of evolution of the Universe settles. The first point of inflection
$a_1$ corresponds to the early stage of the Universe. The second point of
inflection $a_2$ (the end of the Friedmann era and the beginning of the
"second inflation" era) is determined by the condition (\ref{eq:usl2}). This
condition on the basis of the observational data yields the viable range of
the possible forms of the dilaton dark matter equation of state, $p =
\gamma \varepsilon$, from the cold dark matter (with $\gamma = \case23$)
up to the superrigid matter (with $\gamma = 1$). Neither the relativistic
matter ($\gamma =\case13$) nor the dust matter ($\gamma = 0$) does not be
allowed to belong to this range. Therefore the theory predicts the dark
matter to be cold and self-interacting that coincides with the
astrophysical data \cite{Sp-St}.

    It should be emphasized that when the equation of state of the dilaton
dark matter is superrigid ($\gamma = 1$) at the very early stage of the
Universe then the inflation-like stage of the evolution of the Universe
occurs until the equation of state of the dilaton matter will change.

    The hypothesis on the dilaton matter as the dark self-interacting (by
means of the dilaton charge) matter explains, why the dark matter is
detected only as a consequence of gravitational effects and cannot be
discovered by means of the nongravitational interaction with particles
for which the dilaton charge $J = 0$. This hypothesis also leads to the
conclusion that the expansion with acceleration begins when the dark matter
energy density becomes equal by order of magnitude to the vacuum energy
density. This is the answer on the Steinhardt's question:  "why is the
quintessence dominating after 15 billion year and not, say, 1.5 billion
years or 150 billion years" \cite{St}.

\appendix
\section*{Appendix A}
\setcounter{section}{1}
\setcounter{equation}{0}

Let us consider a connected 4-dimensional oriented differentiable manifold
${\cal M}$ equipped with a metric $\breve g$ of index 1, a linear connection
$\Gamma$ and a volume 4-form $\eta$. Then a Weyl--Cartan space $CW_4$ is
defined as the space equipped with a curvature 2-form
${\cal R}^{\alpha}\!_{\beta}$ and a torsion 2-form ${\cal T}^{\alpha}$ with
the metric tensor and the connection 1-form obeying the condition
\footnote{Our notations differ by some details from the notations accepted
in \cite{He:pr}.}
\begin{equation}
- {\rm D} g_{\alpha\beta} = {\cal Q}_{\alpha\beta} = \frac{1}{4}
g_{\alpha\beta}{\cal Q}\; , \qquad {\cal Q}:= g^{\alpha\beta}{\cal Q}
_{\alpha\beta} = Q_{\alpha}\theta^{\alpha}\; ,  \label{eq:Q}
\end{equation}
where ${\cal Q}_{\alpha\beta}$ -- a nonmetricity 1-form, ${\cal Q}$ -- a
Weyl 1-form and ${\rm D} := \rmd + \Gamma\wedge\ldots$ -- the exterior
covariant differential. Here $\theta^\alpha$ ($\alpha = 1,2,3,4$) -- cobasis
of 1-forms of the $CW_4$-space ($\wedge$ -- the exterior product operator).

A curvature 2-form ${\cal R}^{\alpha}\!_{\beta}$ and a torsion 2-form
${\cal T}^{\alpha}$,
\begin{equation}
\fl {\cal R}^{\alpha}\!_{\beta}=\frac{1}{2}R^{\alpha}\!_{\beta\gamma\lambda}
\theta^{\gamma}\wedge\theta^{\lambda}\;, \qquad {\cal T}^{\alpha}=\frac{1}{2}
T^{\alpha}\!_{\beta\gamma}\theta^{\beta} \wedge\theta^{\gamma}\;, \qquad
T^{\alpha}\!_{\beta\gamma} = -2\Gamma^\alpha\!_{[\beta\gamma ]}\;.
\label{eq:302}
\end{equation}
are defined by virtue of the Cartan's structure equations,
\begin{eqnarray}
{\cal R}^{\alpha}\!_{\beta}=\rmd\Gamma^{\alpha}\!_{\beta}+\Gamma^{\alpha}\!_
{\gamma}\wedge\Gamma^{\gamma}\!_{\beta}\;, \qquad {\cal T}^{\alpha} =
{\rm D}\theta^{\alpha} = \rmd\theta^{\alpha}+\Gamma^
{\alpha}\!_{\beta}\wedge\theta^{\beta}\;.  \label{eq:304}
\end{eqnarray}
The Bianchi identities for the curvature 2-form, the torsion 2-form and the
Weyl 1-form are valid \cite{He:pr},
\begin{equation}
{\rm D} {\cal R}^{\alpha}\!_{\beta} = 0\;, \qquad  {\rm D} {\cal T}^{\alpha} =
{\cal R}^{\alpha}\!_{\beta}\wedge \theta^\beta \;, \qquad
\rmd{\cal Q} = 2{\cal R}^{\gamma}\!_{\gamma} \;.  \label{eq:Bian}
\end{equation}

It is convenient to use the auxiliary fields of 3-forms $\eta_\alpha$,
2-forms $\eta_{\alpha\beta}$, 1-forms $\eta_{\alpha\beta\gamma}$ and 0-forms
$\eta_{\alpha\beta\gamma\lambda}$ with the properties,
\begin{eqnarray}
\fl \eta_{\alpha} = \vec{e}_{\alpha}\rfloor \eta = *\theta_{\alpha}\; ,\qquad
&\eta_{\alpha\beta\gamma} = \vec{e}_{\gamma}\rfloor \eta_{\alpha\beta} =
*(\theta_{\alpha}\wedge\theta_{\beta}\wedge \theta_{\gamma})\;,
\label{eq:eta1} \\
\fl \eta_{\alpha\beta} = \vec{e}_{\beta} \rfloor \eta_{\alpha} =
*(\theta_{\alpha}\wedge \theta_{\beta})\;, \qquad
&\eta_{\alpha\beta\gamma\lambda} = \vec{e}_{\lambda}\rfloor
\eta_{\alpha\beta\gamma} = *(\theta_{\alpha}\wedge \theta_{\beta}\wedge
\theta_{\gamma}\wedge \theta_{\lambda}) \;.  \label{eq:eta4}
\end{eqnarray}
Here $*$ is the Hodge operator and $\rfloor$ is the operation of contraction
(interior product) which obeys to the Leibnitz antidifferentiation rule,
\begin{equation}
\vec{v}\rfloor (\Phi\wedge\Psi ) = (\vec{v}\rfloor\Phi )\wedge\Psi + (-1)^p
\Phi \wedge (\vec{v}\rfloor \Psi )\;,  \label{eq:Lrule}
\end{equation}
where $\Phi$ is a $p$-form.
\par
  The properties (\ref{eq:eta1}), (\ref{eq:eta4}) lead to the following
useful relations,
\begin{eqnarray}
\theta^{\sigma} \wedge \eta_{\alpha} = \delta^{\sigma}_{\alpha}\eta \; ,
\qquad \theta^{\sigma} \wedge \eta_{\alpha_1\dots\alpha_p} = (-1)^{p-1}p
\delta^{\sigma}_{[\alpha_1}\eta_{\alpha_2\dots\alpha_p ]}\;,
\label{eq:teta1} \\
\theta^{\sigma} \wedge\theta^\rho \wedge \eta_{\alpha\beta} =
2\delta^{\sigma}_{[\alpha}\delta_{\beta ]}^\rho \eta \;, \label{eq:teta2}\\
\theta^{\sigma} \wedge\theta^\rho \wedge \eta_{\alpha\beta\gamma} =
3\theta^\sigma\wedge \delta^{\rho}_{[\alpha} \eta_{\beta\gamma ]} =
6\delta^{\sigma}_{[\alpha} \delta^\rho_{\beta}\eta_{\gamma ]}\;,
\label{eq:teta3} \\
\theta_{[\alpha}\wedge *(\theta_{\beta ]}\wedge \theta_\gamma ) =
-\frac{1}{2}\eta_{\alpha\beta}\wedge\theta_\gamma \;, \label{eq:teta4}\\
\vec{e}_{\lambda}\rfloor (\theta^{\alpha_1}\wedge\dots\wedge
\theta^{\alpha_p} ) = p\delta^{[\alpha_1}_{\lambda}\theta^{\alpha_2}\dots
\wedge\theta^{\alpha_p ]} \; . \label{eq:svp}
\end{eqnarray}
In space $CW_4$ the equality ${\rm D}\eta = 0$ is fulfilled and the
following formulae are valid \cite{He:pr},
\begin{eqnarray}
\fl {\rm D}\eta_{\alpha\beta\gamma\lambda} = - \frac{1}{2}{\cal Q}
\eta_{\alpha\beta\gamma\lambda}\; , \qquad &{\rm D}\eta_{\alpha\beta\gamma}
= - \frac{1}{2}{\cal Q}\wedge \eta_{\alpha\beta\gamma} + {\cal T}^\lambda
\eta_{\alpha\beta\gamma \lambda}\; ,  \label{eq:Deta1} \\
\fl {\rm D}\eta_{\alpha\beta} = - \frac{1}{2}{\cal Q}\wedge
\eta_{\alpha\beta} + {\cal T}^\gamma \wedge \eta_{\alpha\beta\gamma}\;, \qquad
&{\rm D}\eta_{\alpha} = - \frac{1}{2}{\cal Q}\wedge \eta_{\alpha} +
{\cal T}^ \beta \wedge \eta_{\alpha\beta}\;.  \label{eq:Deta2}
\end{eqnarray}

\section*{Appendix B}
\setcounter{section}{2}
\setcounter{equation}{0}

The variational procedure in the exterior form language is based on the
master formula derived the following Lemma, proved in \cite{bfkl} (see also
\cite{LosH}).

{\em Lemma.} Let $\Phi$ and $\Psi$ be arbitrary $p$-forms defined on $n$
-dimensional manifold. Then the variational identity for the commutator of
the variation operator $\delta$ and the Hodge star operator $*$ is valid,
\begin{eqnarray}
 \Phi \wedge \delta * \Psi = \delta \Psi \wedge * \Phi   \nonumber \\
 + \delta g_{\sigma\rho}\left (\frac{1}{2} g^{\sigma\rho}\Phi\wedge *\Psi
 + (-1)^{p(n-1) + s + 1}\theta^\sigma \wedge * \left (*\Psi\wedge\theta^\rho
\right )\wedge * \Phi \right )  \nonumber \\
 + \delta \theta^\alpha \wedge \left ({(-1)}^p \Phi \wedge * \left (\Psi
\wedge \theta_\alpha \right) + {(-1)}^{p(n-1) + s + 1} * \left
(*\Psi\wedge\theta_\alpha \right ) \wedge * \Phi \right )\,. \label{eq:master}
\end{eqnarray}

The variational procedure is realized with the help of the computation rules
\cite{Lichn},
\begin{eqnarray}
 **\Psi = (-1)^{p(n-p) + s}\Psi \; , \qquad &\Phi\wedge *\Psi = \Psi\wedge
*\Phi \; ,  \label{eq:com1} \\
 {\vec e}_\alpha \rfloor *\Psi = * (\Psi\wedge \theta_\alpha)\;, \qquad
&\theta^{\alpha}\wedge ({\vec e}_\alpha \rfloor \Psi) = p\Psi\;,
\label{eq:com2}
\end{eqnarray}
where $\Psi$ and $\Phi$ are $p$-forms and $s = \mbox{Ind}(\breve g)$ is the
index of the metric $\breve g$, which is equal to the number of negative
eigenvalues of the diagonalized metric. The relations (\ref{eq:com1}) and
(\ref{eq:com2}) lead to the consequences,
\begin{eqnarray}
 {\vec e}_\alpha \rfloor \Psi = (-1)^{n(p-1) + s}
*\!(\theta_{\alpha}\wedge *\Psi )\;,  \label{eq:rul1} \\
 \ast ({\vec e}_\alpha \rfloor \Psi ) = (-1)^{p-1} \theta_{\alpha}\wedge
*\Psi \;,  \label{eq:rul2} \\
 \ast ({\vec e}_\alpha \rfloor *\!\Psi ) = (-1)^{(n-1)(p + 1) + s}
\Psi\wedge \theta_{\alpha}\;.  \label{eq:rul3}
\end{eqnarray}

Let apply the master formula (\ref{eq:master}) to the variation of the
Lagrangian density 4-form (\ref{eq:tlag}), the general relations
(\ref{eq:varr})--(\ref{eq:varq}) and the computation rules
(\ref{eq:teta1})--(\ref{eq:Deta2}), (\ref{eq:com1})--(\ref{eq:rul3}) being
used. The results of the variational procedure for every term of
(\ref{eq:tlag}) have the following form, the exact forms being omitted,
\begin{eqnarray}
2f_0 &:& \delta\Gamma^\alpha\!_\beta \wedge \left (-\frac{1}{4}{\cal Q}\wedge
\eta_\alpha\!^\beta + \frac{1}{2}{\cal T}_\lambda \wedge \eta
_\alpha\!^{\beta\lambda } + \frac{1}{2}\eta_{\alpha\gamma}\wedge
{\cal Q}^{\beta\gamma} \right )   \nonumber \\
&+& \delta g_{\sigma\rho}\left (\frac{1}{2}g^{\sigma\rho} {\cal R}
^\alpha\!_\beta \wedge\eta_\alpha\!^\beta  + \frac{1}{2}\theta^\sigma
\wedge \theta_\beta \wedge *{\cal R}^{\beta\rho}\right ) \nonumber \\
&+& \delta \theta^\sigma \wedge \left (\frac{1}{2}{\cal R}^\alpha\!_\beta
\wedge \eta_\alpha\!^\beta\!_\sigma \right )\;,  \label{eq:lin} \\
2f_0\; \varrho_1 &:& \delta\Gamma^\alpha\!_\beta \wedge 2\theta^\beta \wedge
*{\cal T}_\alpha   \nonumber \\
&+& \delta g_{\sigma\rho}\left ( {\cal T}^\sigma \wedge *{\cal T}^\rho
+ \frac{1}{2} g^{\sigma\rho}{\cal T}_\alpha\wedge *{\cal T}^\alpha
+\theta^\sigma \wedge *(*{\cal T}^\alpha \wedge\theta^\rho ) \wedge
*{\cal T_\alpha}\right )  \nonumber \\
&+& \delta \theta^\sigma \wedge \left (2{\rm D}*{\cal T}_\sigma +
{\cal T}^\alpha \wedge *({\cal T}_\alpha\wedge \theta_\sigma ) +
*(*{\cal T}_\alpha\wedge\theta_\sigma ) \wedge *{\cal T}^\alpha \right ) \;,
\label{eq:ro1} \\
2f_0\; \varrho_2 &:& \delta\Gamma^\alpha\!_\beta \wedge 2\theta^\beta\wedge
\theta_\gamma \wedge *({\cal T}^\gamma\wedge\theta_\alpha )
+ \delta g_{\sigma\rho}\biggl (\frac{1}{2} g^{\sigma\rho} ({\cal T}^\alpha
\wedge \theta_\beta )  \nonumber \\
&+& 2\delta^\sigma_\beta {\cal T}^\alpha \wedge \theta^\rho
- \theta^\sigma \wedge *(*({\cal T}^\alpha \wedge \theta_\beta )
\wedge \theta^\rho )\biggr )\wedge *({\cal T}^\beta \wedge\theta_\alpha )
\nonumber \\
&+& \delta\theta^\sigma \wedge\biggl (2{\rm D} \Bigl (\theta_\alpha\wedge
*({\cal T}^\alpha \wedge \theta_\sigma )\Bigr )
- *({\cal T}^\beta \wedge \theta_\alpha\wedge\theta_\sigma )
({\cal T}^\alpha \wedge \theta_\beta )  \nonumber \\
&+& 2{\cal T}^\alpha \wedge * ({\cal T}_\sigma\wedge \theta_\alpha )
- *\Bigl (*({\cal T}^\beta\wedge\theta_\alpha )\wedge \theta_\sigma\Bigr )
\wedge *({\cal T}^\alpha \wedge\theta_\beta )\biggr ) \;,  \label{eq:ro3} \\
2f_0\; \varrho_3 &:& \delta\Gamma^\alpha\!_\beta \wedge 2\theta^\beta \wedge
\theta_\alpha \wedge *({\cal T}^\gamma\wedge\theta_\gamma )
+ \delta g_{\sigma\rho} \biggl (\frac{1}{2} g^{\sigma\rho} ({\cal T}^\beta
\wedge \theta_\beta )  \nonumber \\
&+& 2{\cal T}^\sigma \wedge \theta^\rho - \theta^\sigma \wedge
*\Bigl (*({\cal T}^\beta \wedge \theta_\beta ) \wedge \theta^\rho \Bigr )
\biggr ) \wedge *({\cal T}^\alpha \wedge\theta_\alpha )  \nonumber\\
&+& \delta\theta^\sigma \wedge \biggl (2{\rm D}\Bigl (\theta_\sigma\wedge
*({\cal T}^\alpha \wedge \theta_\alpha )\Bigr ) - *({\cal T}^\beta \wedge
\theta_\beta \wedge \theta_\sigma ) {\cal T}^\alpha \wedge \theta_\alpha
\nonumber \\
&+& 2{\cal T}_\sigma \wedge * ({\cal T}^\alpha\wedge \theta_\alpha )
- *\Bigl (*({\cal T}^\beta \wedge\theta_\beta )\wedge \theta^\sigma\Bigr )
\wedge *({\cal T}^\alpha \wedge\theta_\alpha )\biggr ) \;, \label{eq:ro2} \\
2f_0\;\lambda &:& \delta\Gamma^\alpha\!_\beta \wedge \left (\frac{1}{2}
\delta^\beta_\alpha {\rm D} *{\cal R}^\gamma\!_\gamma \right )    \nonumber\\
&+& \delta g_{\sigma\rho}\left (\frac{1}{8}g^{\sigma\rho}
{\cal R}^\alpha\!_\alpha \wedge *{\cal R}^\beta\!_\beta
+ \frac{1}{4} \theta^\sigma \wedge *(*{\cal R}^\alpha\!_\alpha \wedge
\theta^\rho)\wedge *{\cal R}^\beta\!_\beta \right )  \nonumber \\
&+& \delta\theta^\sigma \wedge \left (\frac{1}{4} {\cal R}^\beta\!_\beta \wedge
*({\cal R}^\alpha\!_\alpha \wedge\theta_\sigma ) + \frac{1}{4} *(*{\cal R}
^\alpha\!_\alpha \wedge \theta_\sigma) \wedge *{\cal R}^\beta\!_\beta \right )
 \;, \label{eq:lam} \\
2f_0\; \xi &:& \delta\Gamma^\alpha\!_\beta \wedge (4\delta^\beta_\alpha
*{\cal Q} )  \nonumber \\
&+& \delta g_{\sigma\rho}\left (2 g^{\sigma\rho}{\rm D}*{\cal Q} + \frac{1}{2}
g^{\sigma\rho}{\cal Q}\wedge *{\cal Q} - *(*{\cal Q} \wedge \theta^\rho
)\theta^\sigma\wedge *{\cal Q}\right )  \nonumber \\
&+& \delta\theta^\sigma \wedge \left (-{\cal Q}\wedge *({\cal Q}\wedge
\theta_\sigma ) - *(*{\cal Q}\wedge \theta_\sigma ) \wedge *{\cal Q} \right )
\;, \label{eq:xi} \\
2f_0\;\zeta &:& \delta\Gamma^\alpha\!_\beta \wedge \left (2\delta
^\beta_\alpha\theta^\gamma \wedge *{\cal T}_\gamma + \theta^\beta\wedge
*({\cal Q}\wedge \theta_\alpha )\right )  \nonumber \\
&+& \delta g_{\sigma\rho} \biggl (g^{\sigma\rho}{\cal T}^\alpha \wedge
*{\cal T}_\alpha - g^{\sigma\rho} \theta^\alpha \wedge D*{\cal T}_\alpha
+ \frac{1}{2}g^{\sigma\rho}{\cal Q}\wedge \theta^\alpha \wedge
*{\cal T}_\alpha \nonumber \\
&+& \theta^\sigma \wedge *(*{\cal T}_\alpha \wedge \theta^\rho ) \wedge
*({\cal Q}\wedge \theta^\alpha ) + {\cal T}^\sigma \wedge *({\cal Q}\wedge
\theta^\rho ) \biggr )  \nonumber \\
&+& \delta\theta^\sigma \wedge \biggl ({\rm D}*({\cal Q}\wedge\theta_\sigma )
 + {\cal Q}\wedge\theta^\alpha \wedge *({\cal T}_\alpha\wedge\theta_\sigma )
\nonumber \\
&+& *(*{\cal T}_\alpha\wedge\theta_\sigma ) \wedge *({\cal Q} \wedge
\theta^\alpha ) -{\cal Q}\wedge *{\cal T}_\sigma \biggr )\;, \label{eq:zeta}\\
2f_0\;\Lambda &:& \delta g_{\sigma\rho} \left (\frac{1}{2}g^{\sigma\rho}\eta
\right ) + \delta\theta^\sigma\wedge\eta_\sigma \;.  \label{eq:Lam}
\end{eqnarray}
The variation of the term with the Lagrange multiplier in (\ref{eq:tlag})
has the form,
\begin{equation}
\fl \delta\Lambda^{\alpha\beta} \wedge \left ({\cal Q}_{\alpha\beta} -
\frac{1}{4}g_{\alpha\beta}{\cal Q} \right ) + \delta\Gamma^\alpha\!_\beta
\wedge (-2\Lambda_\alpha\!^\beta ) + \delta g_{\sigma\rho}\left (-{\rm D}
\Lambda^{\sigma\rho} - \frac{1}{4}\Lambda^{\sigma\rho}\wedge
{\cal Q} \right ) \; .  \label{eq:Lamab}
\end{equation}

\section*{Appendix C}
\setcounter{section}{3}
\setcounter{equation}{0}

In a Weyl--Cartan space the following decomposition of the connection 1-form
is valid,
\begin{equation}
\Gamma^{\alpha}\!_{\beta} = \stackrel{C}{\Gamma}\!^{\alpha}\!_{\beta} +
\Delta^{\alpha}\!_{\beta}\; , \qquad \Delta_{\alpha\beta} = \frac{1}{8}
(2\theta_{[\alpha}Q_{\beta ]} + g_{\alpha\beta} {\cal Q})\; ,
\label{eq:Grusl}
\end{equation}
where $\stackrel{C}{\Gamma}\!^{\alpha}\!_{\beta}$ denotes a connection
1-form of a Riemann--Cartan space $U_{4}$ with curvature, torsion and a
metric compatible with a connection. This decomposition of the connection
induces corresponding decomposition of the curvature 2-form \cite{bfmod2},
\begin{eqnarray}
\fl {\cal R}^{\alpha}\!_{\beta} = \stackrel{C}{{\cal R}}\!^{\alpha}\!_{\beta}
+ \stackrel{C}{{\rm D}}\Delta^{\alpha}\!_{\beta} +
\Delta^{\alpha}\!_{\gamma}\wedge \Delta^{\gamma}\!_{\beta} = \stackrel{C}
{{\cal R}}\!^{\alpha}\!_{\beta} + \frac{1}{4} \delta^{\alpha}_{\beta}{\cal R}
^{\gamma}\!_{\gamma} + {\cal P}^{\alpha}\!_{\beta}\;,  \label{eq:red1} \\
\fl {\cal P}_{\alpha\beta} = \frac{1}{4}\left ({\cal T}_{[\alpha}Q_{\beta ]}
- \theta_{[\alpha}\wedge \stackrel{C}{{\rm D}}Q_{\beta ]}
+ \frac{1}{8}\theta_{[\alpha}Q_{\beta ]} \wedge {\cal Q} - \frac{1}{16}
\theta_{\alpha} \wedge \theta_{\beta} Q_{\gamma}Q^{\gamma}\right )\; ,
\label{eq:red2}
\end{eqnarray}
where $\stackrel{C}{{\rm D}}$ is the exterior covariant differential with
respect to the Riemann--\-Car\-tan con\-nec\-ti\-on 1-form $\stackrel{C}
{\Gamma}\!^{\alpha}\!_{\beta}$ and $\stackrel{C}{{\cal R}}\!^{\alpha}\!_
{\beta}$ is the Riemann--Cartan curvature 2-form. The decomposition (\ref
{eq:red1}) contains the Weyl segmental curvature 2-form
${\cal R}^{\gamma}\!_{\gamma}$ (\ref{eq:hom}).

    The Riemann--Cartan connection 1-form can be decomposed as follows
\cite{He:pr},
\begin{eqnarray}
\stackrel{C}{\Gamma }\!^{\alpha }\!_{\beta } = \stackrel{R}{\Gamma }
\!^{\alpha }\!_{\beta }+{\cal K}^{\alpha }\!_{\beta }\;, \qquad
{\cal T}^{\alpha } = {\cal K}^{\alpha }\!_{\beta }\wedge \theta ^{\beta }\;,
\label{eq:kon1} \\
{\cal K}_{\alpha \beta } = 2\vec{e}_{[\alpha }\rfloor{\cal T}_{\beta ]}
- \frac{1}{2}\vec{e}_{\alpha }\rfloor \vec{e}_{\beta }
\rfloor ({\cal T}_{\gamma }\wedge \theta ^{\gamma })\;,  \label{eq:kon2}
\end{eqnarray}
where $\stackrel{R}{\Gamma }\!^{\alpha }\!_{\beta }$ is a Riemann
(Levi--Civita) connection 1-form and ${\cal K}^{\alpha }\!_{\beta }$ is a
contortion 1-form.

    The decomposition (\ref{eq:kon1}) of the connection induces the
decomposition of the cur\-va\-tu\-re as follows,
\begin{equation}
\stackrel{C}{{\cal R}}\!^{\alpha}\!_{\beta} = \stackrel{R}{{\cal R}}
\!^{\alpha}\!_{\beta} + \stackrel{R}{{\rm D}}{\cal K}^{\alpha}\!_{\beta} +
{\cal K}^{\alpha}\!_{\gamma}\wedge {\cal K}^{\gamma}\!_{\beta}\;,
\label{eq:red3}
\end{equation}
where $\stackrel{R}{{\rm R}}\!^{\alpha}\!_{\beta}$ is the Riemann curvature
2-form and $\stackrel{R}{{\rm D}}$ is the exterior covariant differential
with respect to the Riemann connection 1-form $\stackrel{R}{\Gamma}\!^
{\alpha}\!_{\beta}$.

    Let us substitute the decomposition (\ref{eq:red2}) into the equation
(\ref{eq:ein}) and after this use the decomposition (\ref{eq:red3}), the
relation (\ref{eq:T}) being taken into account. We get the following results
for the every term of the equation (\ref{eq:ein}). The linear term or this
equation decomposes as follows,
\begin{eqnarray}
\fl - \left (\stackrel{R}R\!^{\alpha}\!_{\sigma} - \frac{1}{2}
\delta^\alpha_\sigma\stackrel{R}R \right )\eta_\alpha + \frac{2}{3}
\left (\stackrel{R}\nabla_\alpha T^\alpha \right )\eta_\sigma - \frac{2}{3}
\left (\stackrel{R}\nabla_\sigma T^\alpha \right ) \eta_\alpha  - \frac{1}{9}
T_\alpha T^\alpha \eta_\sigma - \frac{2}{9}T_\sigma T^\alpha \eta_\alpha
\nonumber \\
+ \frac{1}{12} T_\alpha Q^\alpha \eta_\sigma
+ \frac{1}{12} T_\sigma Q^\alpha\eta_\alpha
+ \frac{1}{12}Q_\sigma T^\alpha \eta_\alpha  \nonumber \\
+ \frac{1}{4}\left (\stackrel{R}\nabla_\sigma Q^\alpha \right ) \eta_\alpha -
\frac{1}{4} \left (\stackrel{R}\nabla_\alpha Q^\alpha \right ) \eta_\sigma
- \frac{1}{64}Q_\alpha Q^\alpha \eta_\sigma
- \frac{1}{32} Q_\sigma Q^\alpha \eta_\alpha\;. \label{eq:f1}
\end{eqnarray}
The other terms decompose as follows,
\begin{eqnarray}
\varrho_1 &:& \frac{2}{3}\left (\stackrel{R}\nabla_\sigma T^\alpha \right )
\eta_\alpha - \frac{2}{3}\left (\stackrel{R}\nabla_\alpha T^\alpha \right )
\eta_\sigma + \frac{2}{9}T_\sigma T^\alpha \eta_\alpha +
\frac{1}{9}T_\alpha T^\alpha \eta_\sigma \nonumber \\
&&  - \frac{1}{4}Q_\sigma T^\alpha \eta_\alpha - \frac{1}{12}T_\sigma
Q^\alpha \eta_\alpha  + \frac{1}{12} Q_\alpha T^\alpha \eta_\sigma\;,
\label{eq:f2} \\
\varrho_2 &:& - \frac{4}{3}\left (\stackrel{R}\nabla_\sigma T^\alpha \right )
\eta_\alpha + \frac{4}{3}\left (\stackrel{R}\nabla_\alpha T^\alpha \right )
\eta_\sigma - \frac{4}{9}T_\sigma T^\alpha \eta_\alpha - \frac{2}{9}T_\alpha
T^\alpha \eta_\sigma \nonumber \\
&&  + \frac{1}{2}Q_\sigma T^\alpha \eta_\alpha + \frac{1}{6}T_\sigma
Q^\alpha \eta_\alpha  - \frac{1}{6} Q_\alpha T^\alpha \eta_\sigma\;,
\label{eq:f3} \\
\zeta &:& - \left (\stackrel{R}\nabla_\alpha Q^\alpha \right ) \eta_\sigma +
\left (\stackrel{R}\nabla_\sigma Q^\alpha \right ) \eta_\alpha +
\frac{2}{3}T_\alpha Q^\alpha \eta_\sigma \nonumber \\
&& - \frac{2}{3}Q_\sigma T^\alpha \eta_\alpha + \frac{1}{8}Q_\alpha Q^\alpha
\eta_\sigma - \frac{1}{2} Q_\sigma Q^\alpha \eta_\alpha\;, \label{eq:f4}\\
\xi &:& Q_\alpha Q^\alpha \eta_\sigma - 2 Q_\sigma Q^\alpha \eta_\alpha\;.
\label{eq:f5}
\end{eqnarray}

\end{document}